\documentclass[twocolumn,showpacs,prb,amsmath,amssymb,floatfix,eqsecnum]{revtex4}
\usepackage{dcolumn}
\usepackage{amsmath,amssymb}
\usepackage{bm}
\usepackage{epsfig}

\newcommand{\bd}{\bm}

\begin{document}

\title{Functional renormalization group approach to interacting bosons at zero temperature}

\author{Andreas Sinner$^1,$ Nils Hasselmann,$^2$ and Peter Kopietz$^3$}

\vspace{2mm}  

\affiliation{$^1$Institut f\"ur Physik, Theorie II, Universit\"at 
Augsburg, Universit\"atsstra\ss e 1, 86159 Augsburg, Germany \\
%\address{
$^2$International Institute of Physics, Universidade Federal do Rio
Grande do Norte, 59072-970 Natal/RN, Brazil \\
%\address{
$^3$Institut f\"{u}r Theoretische Physik, Universit\"{a}t
    Frankfurt, Max-von-Laue-Stra\ss e 1, 60438 Frankfurt, Germany}
%\ead{sinner@itp.uni-frankfurt.de, hasselma@itp.uni-frankfurt.de, pk@itp.uni-frankfurt.de}

%\date{\today}
\date{August 25, 2010}
%\date{March XX, 2005}

\begin{abstract}
\noindent
We investigate the single-particle spectral density of interacting bosons 
within the non-perturbative
functional renormalization group technique. 
The flow equations 
for a Bose gas 
are derived in a scheme which treats the
two-particle density-density correlations exactly but neglects irreducible
correlations among three and more particles.
These flow equations are solved within 
a truncation which allows to extract the complete frequency and momentum structure of
the normal and anomalous self-energies. Both the asymptotic
small momentum regime, where perturbation regime fails,
as well as the perturbative regime at larger
momenta are well described within a single unified approach.
%The non-selfconsistent solutions builds
%on a derivative expansion which captures the non-perturbative aspects of the
%interacting Bose gas as.
 The self-energies do not exhibit any infrared divergences,
satisfy the $U(1)$ symmetry constraints,  and 
are in accordance with the Nepomnyashchy relation
which states that the anomalous self-energy vanishes at zero
momentum and zero frequency.
From the self-energies we extract the single-particle spectral
density of the two-dimensional Bose gas. 
The dispersion is found to be of the Bogoliubov form and
shows the crossover from linear Goldstone modes to 
the quadratic behavior of  quasi-free bosons. The damping of the quasiparticles
is found to be in accordance with the standard Beliaev damping. 
We furthermore 
recover the exact asymptotic limit of the propagators derived
by Gavoret and Nozi\`{e}res
and discuss the nature of the non-analyticities of the self-energies in the
very small momentum regime. 

% We study systems of interacting bosons with condensate at zero temperature within the framework of the Functional Renormalization Group (FRG). We derive flow equations for coupling parameters appearing in the expansion of the effective action up to the second order in spatial and temporal variables and investigate them numerically. We discuss the role of the generalized Ginzburg scale for the RG-flow of the effective action and extract this scale from the flow equation for the interaction. Our equations describe correctly the limit of hard-core bosons. The approach enables us to calculate momentum dependent Matsubara Green's functions and from them the single-particle spectral function. This paper represents an extended version of our recent Phys. Rev. Lett.~[\onlinecite{SHK2009}].
\end{abstract}

\maketitle
\section{Introduction}
\label{sec:Intro}

Interacting Bose gases in the continuum have been studied 
with various approaches for more than half a century, yet a
controlled approach exists only for the weakly interacting 
variety and even there the asymptotic regime at small frequencies
and momenta proved to be non-perturbative.
The first qualitatively correct description of the excitation spectrum 
of weakly interacting bosons was 
given by Bogoliubov \cite{Bogoliubov1947} using a mean-field 
approximation. The Bogoliubov excitation spectrum exhibits the gapless 
linear Goldstone mode character at small wave vectors and approaches the 
quadratic dispersion of free bosons as wave vectors become large and 
interaction effects become negligible. 
%It is characterized by a single
The crossover scale separating these two regimes is given by
%The single scale determining the Bogoliubov spectrum is the
%crossover scale %between the two regimes is
$k_c=2mc$ where $m$ is the mass of the bosons and $c$ the
velocity of the Goldstone modes.
In dimension $D=3$, this picture has recently 
been confirmed experimentally using the Bragg spectroscopy technique 
on cold atoms.\cite{Steinhauer2002} 

Beliaev \cite{Beliaev58} 
went beyond the 
Bogoliubov approximation %, correct to leading order in the interaction,
and calculated the self-energies of the quasiparticles to second
order in the effective interaction, allowing  both to extract
corrections to the quasiparticle dispersion as well as to calculate
their lifetime. 
%methods, 
%were applied to systems of weakly interacting bosons very soon 
%afterwards \cite{Beliaev58}. They allowed not only to correctly 
%reproduce Bogoliubov results but also to calculate the leading order 
%corrections to the excitation spectrum and dynamical quantities like the 
%damping of quasi-particles. 
However, as discussed in detail in Ref.~[\onlinecite{Shi1997}], 
perturbative
approaches suffer from infrared (IR) divergences for all $D\leq 3$
which are difficult
to control at second order (in most physical quantities the divergences
cancel out)
and essentially impossible to control at  any higher order of perturbation theory. 
These divergences 
 have their origin in the $U(1)$ symmetry of
the model and can be traced to the divergence of the longitudinal
correlation 
function.\cite{Nepomnyashchy1975,Pistolesi2004,Vaks68,Patashinskii73,Zwerger04} 
%Very soon after Beliaev work Hugenholtz and Pines 
%[\onlinecite{Hugenholtz1959}] reviewed his results and derived an equality 
%which is now commonly referred as Hugenholtz-Pines equality and relates 
%both normal and anomalous self-energies in the long wavelength limit to 
%he chemical potential. This equality was shown to hold to all orders 
%of the perturbative expansion. The Hugenholtz-Pines relation was later r
%evisited by 
Nepomnyashchy and Nepomnyashchy \cite{Nepomnyashchy1975} 
showed rigorously that the IR divergences, together with 
Ward identities originating from the $U(1)$ symmetry of the model,
lead to the surprising result that the anomalous self-energy 
$\Sigma^A({\bd k},\omega)$ vanishes
at zero momentum and frequency. This result cannot be reproduced by 
finite order perturbation theory, and is in sharp contrast
with both the Bogoliubov and the Beliaev approach where the
anomalous self-energy at zero momentum and zero frequency is finite. 
Furthermore,  in the perturbative approaches
a finite value of  $\Sigma^A(0)$ is a key
quantity which controls both the crossover to the Goldstone regime
as well as the velocity of the Goldstone modes. Nepomnyashchy and 
Nepomnyashchy showed further that
the correct 
structure of the self-energies contains 
non-analytic terms 
which are dominant in the asymptotic limit of small frequencies and
momenta. This  non-perturbative structure arises for modes with
momenta smaller than the generalized Ginzburg scale
$k_G$, which, for weakly interacting bosons, is much smaller than
$k_c= 2 m c$. 
A correct analysis of these non-analytic terms allows to recover
the Goldstone modes within a framework where $\Sigma^A(0)=0$.
There are a number of approaches which avoid the IR divergences,
the earliest is due to Popov who separated the low energy modes
from the high energy modes and treated the low energy sector within
a hydrodynamic theory for the phase degrees of freedom, which is
free of IR divergences.\cite{Popov79} Another approach is to
artificially break the $U(1)$ symmetry by a small symmetry breaking
term which can in the end be removed to yield finite results
for all physical observables.\cite{Capogrosso2010} A natural
framework to deal with IR divergences is the renormalization group (RG).
A field theoretical RG analysis of the interacting Bose gas 
at asymptotically small frequencies and momenta 
has been developed in Ref.~[\onlinecite{Pistolesi2004}], which recovers
the asymptotic behavior of the self-energies and the single particle
Green's functions. This required however
a very careful analysis of Ward identities.

In several recent publications the non-perturbative or 
functional renormalization group (FRG) has been
applied to the interacting Bose gas.
\cite{Dupuis2007,Wetterich2007,SHK2009,Dupuis2009,Dupuis2009b,Eichler09}
The FRG %is a non-perturbative RG technique 
%which 
is based on an exact flow equation of the generating functional
of irreducible vertex functions (average effective action).\cite{Wetterich93}
Several approaches were based on a derivative expansion of the
effective action which is sufficient to find a relatively simple
description of the asymptotic regime.\cite{Wetterich2007,Dupuis2007} 
One advantage
of the FRG approach is that the Ward identities associated with the
$U(1)$ symmetry are automatically obeyed if the truncation 
does not violate the invariance of
the effective action. Another advantage of the FRG is
that it can be employed to study the full momentum and frequency
dependence of the self-energies.
See Refs.~[\onlinecite{Ledowski04,Blaizot06,SHK2008}]
for FRG calculations of the momentum dependent self-energy of classical models.
In Ref.~[\onlinecite{SHK2009}] we showed how the FRG can be 
used to calculate the spectral density of the interacting
Bose gas. This was based on a truncation of the exact FRG
flow equation which 
respects the Nepomnyashchy relation and recovers the non-analytic
structure of the low energy sector. Moreover, our FRG approach also  
describes the full dispersion of the Bogoliubov spectrum as well
as the Beliaev damping of the quasiparticles. 
This is the first approach which
succeeds to describe both the non-perturbative aspects which 
appear at the very small wavevector scale $k\ll k_G$, 
as well as the perturbative
ones, which are controlled by the larger scale 
$k_c=2m c$, %which are controlled by the scale $k_c$,
within a unified framework. Recently Dupuis \cite{Dupuis2009,Dupuis2009b} 
developed a similar approach
and clarified the role of $k_G$ in the flow equations, but
did not resolve the Beliaev damping of the quasiparticles.

Here we present in detail our approach of Ref.~[\onlinecite{SHK2009}]
and investigate analytically  the different scaling regimes of the
flow equations. We
further derive the flow equations for interacting Bose systems 
which treat the irreducible density-density interactions exactly
and which has not been published previously. This set of equations,
while difficult to solve, would potentially allow to analyze
strongly interacting yet dilute Bose systems, which can now 
be realized in experiments and whose spectrum shows clear deviations
from the prediction of the Beliaev approach.\cite{Papp08}

This work consists of the following parts: 
In Sec.~\ref{sec:nepo} 
we define the model and briefly review the structure of the one-particle
Green's functions, i.e. discuss the excitations and their damping. We further
summarize the structure of the self-energies in the non-perturbative regime.
In Sec.~\ref{sec:FRG} we discuss the FRG approach, introduce our approximation
and discuss both the general structure of the flow equations as well
as the results of a derivative expansion in subsection 
\ref{sec:derivative} for which we present results in $D=2$. 
In Sec.~\ref{sec:Spectral} we finally discuss the
results of a non-self-consistent solution of the flow equations which
yields the full momentum and frequency dependence of the self-energies
and allows to extract the single-particle spectral function. Numerical
results are shown for $D=2$. 
In Appendix~\ref{app:Nepo} we briefly review the arguments from
Ref.~[\onlinecite{Nepomnyashchy1975}] which lead to $\Sigma^A(0)=0$.
An analysis of the derivative expansion is presented 
in Appendix~\ref{sec:HardCore}, where the flow of the
effective interaction is analyzed analytically and the standard
$T$-matrix results from diagrammatic approaches are recovered for general
dimension $D$. Furthermore, in Appendix~\ref{sec:GizbScale} we show how
the Ginzburg scale $k_G$ emerges from the flow equations. 
In Appendix~\ref{sec:AsymGreen} we further show analytically, how the 
exact results for the asymptotic form of the normal and anomalous 
propagators is obtained from the non-self-consistent approach of
Sec.~\ref{sec:Spectral}.
We 
conclude our work with a summary in Sec.~\ref{sec:summary}.

% standard Beliaev approach in the functional integral framework. In Section~\ref{sec:FRGformal} we propose the truncation scheme for the effective action and derive RG flow equations for the condensate, for both normal and anomalous self-energy and coupling parameters of the derivative expansion. These equations are studied in Section~\ref{sec:FRGconcret} numerically. In the same section we analyze the velocity of the Goldstone-mode in $D=2$. The approximate analytical solution of the flow-equation for the interaction is obtained in Section~\ref{sec:GizbScale}. In the same section we  obtain an analytical expression for the crossover scale between the quasifree and Goldstone regimes. The hard-core behavior of bosons is examined in Section~\ref{sec:HardCore}. We rederive the Fisher result for the velocity of the Goldstone-mode~[\onlinecite{Fisher1988}] as well as the crossover scale from the quasifree to the hard core regime~[\onlinecite{Schick1971}]. In Section~\ref{sec:AsymGreen} we derive the asymptotic formulas for both normal and anomalous Matsubara Green's functions first obtained by Gavoret and Nozi\`{e}res~[\onlinecite{Gavoret1964}]. These expressions are compared with the numerically obtained propagators in Section~\ref{sec:Spectral}, and are in an excellent accordance with them. Finally, we calculate and discuss the single-particle spectral function and extract from this the spectrum of elementary excitations and damping of quasi-particles.
\section{Self-energies of the interacting Bose gas}
\label{sec:nepo}
%In this section we want to briefly summarize the exact results for the normal
%and anomalous self energies which where obtained in Refs.~[\onlinecite{1975}] by
%Nepomnyashchy and Nepomnyashchy. The key steps on how these results were obtained
%are summarized in Appendix~\cite{app:nepo}. 

We consider a system of bosons at zero temperature  
with a repulsive interaction potential 
$u^{}_{\Lambda^{}_0}(k)$. The bare microscopic action is
defined in terms of the following functional of a complex
field $\psi$,
\begin{eqnarray}
\nonumber
S[\bar\psi,\psi] & = & -\intop_{K} \bar{\psi}^{}_K (i\omega-\epsilon^{}_{k}+\mu)\psi^{}_K \\
\label{eq:InAct}
&& + \frac{1}{2}\intop_{K}u^{}_{\Lambda^{}_0}(k) \rho^{}_{-K} \rho^{}_{K},
\end{eqnarray}
where $K=(\bm{k},i\omega)$ and $\bm{k}$ is a $D$-dimensional 
momentum variable with the absolute value $k$ and $i\omega$ a bosonic Matsubara frequency. 
At zero temperature
the integration  over $K$ is defined as
\begin{equation}
 \intop^{}_{K} \equiv \intop\frac{d\omega}{2\pi} \intop\frac{d^D k}{(2\pi)^D}
\, .
\end{equation}
The first term of Eq.~(\ref{eq:InAct}) describes non-interacting bosons 
with mass $m$ and the dispersion $\epsilon_k=k^2/2m$. The chemical potential
is denoted by
$\mu$.
The second part represents the interaction
where $\rho^{}_K$ denotes the Fourier transform of the local density 
\begin{equation}
\label{eq:DensOp}
\rho_K = \intop_{Q}\bar\psi_{Q}\psi_{Q+K} \, .
\end{equation}
The subscript $\Lambda_0$ of $u^{}_{\Lambda^{}_0}(k)$
indicates that the theory is
assumed to be regularized in the ultraviolet (UV) by a finite
cutoff $\Lambda_0$. This cutoff is related to the finite
range of the interaction; for example, for hard core interactions 
$\Lambda_0\sim 2\pi/d$, where $d$ is the
diameter of the particles. 
%The interaction $u_{\Lambda^{}_0}(K)$ is supposed to be  regularized by 
%means of the ultraviolet (UV) cutoff $\Lambda_0$, i.e. 
%\begin{equation}
%u^{}_{\Lambda^{}_0}(K) = \left\{
%\begin{array}{ccc}
%u^{}_0  & {\rm if} & |{\bm k}|\leqslant\Lambda^{}_0;\\
%0  &   & {\rm else}.
%\end{array}
%\right.
%\end{equation}
The $U(1)$ symmetry of the model Eq.~(\ref{eq:InAct}) is spontaneously
broken in the ground state and 
%with particle conservation  
%is broken spontaneously and
the field $\psi$ acquires a finite grand canonical expectation value, 
\begin{subequations}
\begin{eqnarray}
\label{eq:PsiShift_1} 
\phi^0 = \big< \psi({\bm r},\tau) \big> \neq 0 \, .
\label{eq:PsiShift_2} 
\end{eqnarray}
\end{subequations}
%The expectation value 
%$\big< \psi_K \big>=\delta_{K,0} \phi^0$ is finite and
A macroscopic number of particles in the ground state 
forms a condensate with density $\rho^0=|\phi^0|^2$.
%For notational convenience, we
%shall here assume a real valued $\phi^0$. 
Because $\phi^0\neq 0$,
the Green's function has both normal and anomalous components. 
%For a real
%valued order parameter $\phi^0$, 
The
Dyson equation for the Green's functions in the symmetry broken state can be 
compactly written in a 2$\times$2 matrix in
the space spanned by the field types $\psi_K$ and 
$\bar{\psi}_K$
%matrix form using  a ($\psi_K$, $\bar{\psi}_K$) basis 
as 
\begin{equation}
{\bm G}^{-1}(K) = {\bm G}^{-1}_0(K)-{\bm \Sigma}(K) \, ,
\end{equation}
with 
\begin{eqnarray}
  {\bm G}(K) &=&\left(
    \begin{array}{cc}
      % G_0^{-1}(K)-\Sigma^{N}_\Lambda(K) & -\Sigma^{A}_\Lambda(K) \\
%    -\Sigma^{A}_\Lambda(K) &  G_0^{-1}(K)-\Sigma^{N}_\Lambda(K)
      G^N(K) & G^A(K) \\
      G^A(K)^* & G^N(-K)
    \end{array}
  \right) \, ,
  \\
  {\bm \Sigma}(K)&=&\left(
    \begin{array}{cc}
      \Sigma^N(K) & \Sigma^A(K) \\
      \Sigma^A(K)^* & \Sigma^N(-K)
    \end{array}
  \right)\, ,
 \\
  {\bm G}_{0}^{-1}(K)&=&\left(
    \begin{array}{cc}
      G_{0}^{-1}(K) & 0\\
      0 & G_{0}^{-1}(-K)
    \end{array}
  \right)\, ,
\end{eqnarray}
and 
\begin{equation}
G_0^{-1}(K)=i \omega -\epsilon_k +\mu \, . 
\end{equation}
This leads to 
\begin{eqnarray}
	&&{\bm G}(K) = 
%	\left( 
%	\begin{array}{cc}
%		G^{A}(K) & G^{N}(K)  \\
%		G^{N}(-K) & G^{A}(-K) 
%	\end{array}
%	\right)
	\\ \nonumber
	&& 
        \hspace{-.4cm} 
	\frac{1}{{\cal D}(K)}\left(
	\begin{array}{cc}
          -G^{-1}_0(-K)+\Sigma^{N}(-K) &	\Sigma^{A}(K)  \\
		\Sigma^{A}(K)^* &	-G^{-1}_0(K)+\Sigma^{N}(K)  
	\end{array}
	\right),
	\label{eq:FullProp} 
\end{eqnarray}
where  the denominator ${\cal D}(K)$ is given by 
\begin{eqnarray}
	\nonumber
	{\cal D}(K) &=& - [ G_0^{-1} ( K ) - \Sigma^N ( K )][ G_0^{-1} ( -K ) - 
 \Sigma^N ( -K )]
%[i\omega+\epsilon_k-\mu+\Sigma^{N}(-K)][i\omega-\epsilon_k+\mu-\Sigma^{N}(K)]
\\
	\label{eq:Denom1} 
	&&+|\Sigma^{A}(K)|^2.  
\end{eqnarray}
We shall refer to $G^N(K)$ as the normal propagator and
$G^A(K)$ as the anomalous propagator.\cite{Morawetz10}
The $U(1)$ symmetry imposes constraints on the vertices 
among which the 
Hugenholtz-Pines relation for the self-energies \cite{Hugenholtz1959}
\begin{equation}
\label{eq:HPR}
\Sigma^N(0)-\Sigma^A(0)=\mu
\end{equation}
is the best known. This equation is obeyed by the Bogoliubov approach \cite{Bogoliubov1947}
which approximates both self-energies by the frequency independent
leading order result 
$\Sigma^{N}(K) \approx \rho^0 [u_{\Lambda_0}(0)+u_{\Lambda_0}(k)]$ and 
$\Sigma^{A}_{\Lambda_0}(K) \approx \rho^0 u_{\Lambda_0}(k)$, while the
condensate density is given by $\rho^0 \approx \mu/u_{\Lambda_0}(0)$.  
This yields undamped excitations with dispersion
\begin{equation}
	E_k=\sqrt{\epsilon_k^2 +2\rho^0 u^{}_{\Lambda_0}(k) \epsilon^{}_k} \, .
\end{equation}
For small $k$ the dispersion has a Goldstone mode character with $E_k \sim c_0 k$ with
the velocity 
\begin{eqnarray}
c_0=\sqrt{\frac{\rho^0 u_{\Lambda_0}(0)}{m}},
\end{eqnarray}
while at large momenta $E_k$ approaches the free particle dispersion $\epsilon_k = k^2 /2m$.
If $u_{\Lambda_0}(k)=u_{\Lambda_0}$ is independent of $k$, we can write
\begin{equation}
\label{eq:BogII} 
E_k=c_0 k\sqrt{1+k^2/k^2_{0}},
\end{equation}
where 
 \begin{equation}
 k_{0}=2mc_0
\end{equation}
 is the mean-field estimate for the characteristic scale $k_c = 2 m c$ of the crossover 
from Goldstone modes to quasi-free bosons. 

The leading (second order in the interaction) many-body corrections to the
Bogoliubov approximation have been calculated by
Beliaev.~\cite{Beliaev58}
At this level of approximation the single-particle excitations are damped.
For $k\to 0$ the damping $\gamma_k$ has in three dimensions
the form \cite{Beliaev58,Shi1997}
\begin{equation}
  \label{eq:BelDamp3D}
  \gamma^{(D=3)}_{k} = \frac{3k^5}{640 \pi m \rho^0} \, . 
\end{equation}
while in two dimensions~\cite{Kreisel2008,Chung2008}
\begin{equation}
  \label{eq:BelDamp2D}
  \gamma^{(D=2)}_{k} = \frac{\sqrt{3}c^{}_0 k^3}{32 \pi \rho^0} \, .
\end{equation}
The general result in $D$ dimensions can be written as \cite{Kreisel2008} 
\begin{equation}
  \label{eq:BelDampGen}
  \gamma^{(D)}_{k} \approx \alpha^{}_0 k^{2D-1},
\end{equation}
where
\begin{equation}
\alpha_0= 3^{\frac{D+1}{2}} K_{D-1} \frac{k^{3-D}_0}{32m\rho^0} 
\intop_0^1 dx \, x^{D-1}(1-x)^{D-1} \, ,
\end{equation}
with
\begin{equation}
\label{def:Kd}
K_D=\frac{1}{2^{D-1}\pi^{D/2}\Gamma\left[D/2\right]} \, .
\end{equation}
However, as discussed in detail in Ref.~[\onlinecite{Shi1997}], already
at second order the diagrams for the self-energy are in fact IR
divergent and the divergent terms have to be separated from the non-divergent
terms in order to obtain physical results. These divergences are in fact
related to another problem common to all perturbative approaches, which
is that they violate an exact result obtained by Nepomnyashchy and 
Nepomnyashchy, \cite{Nepomnyashchy1975} who showed that
 the momentum and frequency
independent part of the anomalous self-energy vanishes for $D\leq 3$,
\begin{equation}
  \Sigma^A(K=0)=0 \, .
\end{equation}
This result is a direct consequence of the IR divergences and a 
Ward identity relating the three point vertices to the two point
vertices, see Appendix~\ref{app:Nepo}. 
While in the Bogoliubov and Beliaev approach
a finite value of $\Sigma^A(K=0)$ ensures the Goldstone character of 
the long wavelength modes, in the exact theory the Goldstone
modes are recovered by a non-analytic structure of the 
self-energies which have at small frequencies and momenta 
the form  \cite{Nepomnyashchy1975}
\begin{subequations}
  \begin{eqnarray}
    \label{eq:nepomn:SE}
    \Sigma^N(K) &\approx& \rho^0 u(K)+ \mu+i\omega+
    a_N\epsilon_{k}+b_N\omega^2 \; , \, \, \, \, \, \, \,
    \\
    \label{eq:nepomn:NSE}
    \Sigma^A(K) &\approx& \rho^0 u(K)+a_A\epsilon_k+b_A\omega^2 \; , \, \, \,
  \end{eqnarray}
\end{subequations}
where $u(K)$ is a non-analytic function of $\omega$ and $k$
and the coefficients $a_N,b_N,a_A,b_A$ depend on the
interaction (note that the coefficient of the $i \omega$ term
is exactly equal to unity \cite{Nepomnyashchy1975}). Thus, while each of the self-energies is non-analytic
for $D\leq 3$, their difference is analytic, so that the quantity 
\begin{equation}
\sigma(K)=\Sigma^N(K)-\Sigma^A(K)-\mu
\end{equation}
has the expansion
\begin{equation}
  \label{eq:nepomn:diff} 
  \sigma(K)\approx i\omega+(a_N-a_A)\epsilon_k+(b_N-b_A)\omega^2 \, .
%  + {\cal O}(\omega^2,k^2)\, ,
\end{equation}
We will use in our FRG approach below a scheme where this structure
of the self-energies appears naturally. 
%Moreover, our approach shows
%that, 
As long as all irreducible correlations 
beyond density-density correlations
can be neglected, $u(K)$ is in fact the fully renormalized 
density-density
interaction which also completely determines the anomalous self-energy,
i.e. $\Sigma^A(K)=\rho^0 u(K)$.

\section{Functional renormalization group approach to the interacting Bose gas}
\label{sec:FRG}
% In order to parameterize the model Eq.~(\ref{eq:InAct}), it is convenient to 
% introduce the dimensionless quantities
% \begin{equation}
% \label{eq:diffs:DMLmu}
% \tilde\mu^{}_0~=~2m\mu\Lambda^{-2}_0 = \frac{k^2_\mu}{\Lambda^{2}_0},
% \end{equation} 
% which is the chemical potential $\mu$ in units of the cutoff energy 
% $\Lambda^2_0/2m$. We also write it above in terms of
% the chemical potential related momentum 
% $k^{}_\mu~=~\sqrt{2m\mu}$, which is introduced in analogy to the 
% Fermi-momentum of fermionic systems. Sometimes, the momentum $k^{}_\mu$ is referred to as the inverse healing length. In the theory of Bose-Einstein condensation, the healing length defines the distance over which the condensate wave-function tends to its bulk value when subject to a localized perturbation~[\onlinecite{Pethick}]. Furthermore, we define the dimensionless interaction
% \begin{equation}
% \label{eq:diffs:DMLint}
% \tilde u^{}_{0}(K)= 2m k^{D-2}_\mu u^{}_{\Lambda^{}_0}(K),
% \end{equation}
% which measures the bare interaction in units of the momentum $k^{}_\mu$.
We shall now set up the FRG equations for the interacting
Bose gas. For recent reviews of the FRG method 
see Refs.~[\onlinecite{Morris98,Bagnuls01,Berges02,Delamotte07,Pawlowski07,PKbook,Rosten10}]. 
The basic idea is simple and consists of introducing an IR 
cutoff $\Lambda$ to regulate the theory which is then sent to zero in 
infinitesimally small steps. The FRG
follows the evolution of the various vertex functions
as the IR cutoff $\Lambda$ is lowered
and yields the true vertex functions
for $\Lambda\to 0$.
This will allow us to extract the excitation spectrum and the
damping of quasiparticles from the renormalized self-energies
within a theory which is free of IR
divergences.

%Since a straightforward perturbative calculation of the self-energies in the 
%symmetry broken state,
%starting from Eq.~(\ref{eq:InAct}), leads at second order in the 
%interaction and at zero temperature
%to IR divergences for 
%$D\leq 3$,\cite{Beliaev58,Shi1997} 
The IR divergences which plague the perturbative approaches \cite{Beliaev58,Shi1997} 
are removed if we regulate the 
theory by introducing
a momentum cutoff $\Lambda$ in the free propagator, %by defining 
\begin{equation}
G_{0,\Lambda}^{-1}(K)=i\omega - \epsilon_{k}+\mu - R_\Lambda(k)\, . 
\end{equation}
Here, $R_\Lambda(k)$ is a regulator function which removes
the IR divergences arising from modes with $k<\Lambda$ and
will be specified later.
The FRG approach is based on the cutoff-dependent effective action which
is defined by
\begin{equation}
\Gamma_\Lambda[\bar{\phi},\phi]={\cal L}_\Lambda [\bar{\phi}, \phi]-\intop_K \bar{\phi}_K 
G^{-1}_{0,\Lambda}(K)\phi_K \, ,
\end{equation}
where ${\cal L}_\Lambda [\bar{\phi}, \phi] $ is the cutoff-dependent Legendre transform 
of the generating functional
of connected Green's functions.
The effective action is the generating functional of irreducible vertex 
functions, i.e.
it generates cutoff-dependent  irreducible vertices 
$\Gamma^{ (n,m)}_{\Lambda} ( K_1^{\prime} , \ldots , K^{\prime}_n ; 
K_m, \ldots , K_1 )$
which in the condensed phase are defined via the functional Taylor expansion
of the corresponding  generating functional  in powers
of the fluctuations $\delta \phi_K = \phi_K - \delta_{ K,0} \phi_\Lambda^0$,
\cite{Schuetz2005}
\begin{eqnarray}
  \Gamma_{\Lambda} [ \bar \phi ,  \phi  ]
  & = & \sum_{ n ,m =0}^{\infty} \frac{1}{ n! m!} \intop_{ K_1^{\prime} }
  \cdots \intop_{ K_n^{\prime} } \intop_{ K_m }
  \cdots \intop_{ K_1 } 
  \nonumber
  \\
  & \times & \delta_{ K_1^{\prime} + \ldots + K_n^{\prime} , K_m + \ldots + K_1 }
  \nonumber
  \\
  & \times &
  \Gamma^{ (n,m)}_{\Lambda} ( K_1^{\prime} , \ldots , K^{\prime}_n ; 
  K_m, \ldots , K_1 )
  \nonumber
  \\
  & \times & 
  \delta \bar{\phi}_{ K_1^{\prime}} \ldots  \delta \bar{\phi}_{K_n^{\prime}}
  \delta {\phi}_{ K_m} \ldots  \delta {\phi}_{K_1} \, .
  \label{eq:gendef}
\end{eqnarray}
We normalize $\Gamma_\Lambda [\bar{\phi},\phi]$ such that, 
at the initial RG scale $\Lambda_0$, it reduces to the bare interaction
minus the chemical potential term 
$\mu \intop_K \bar{\phi}_K\phi_K$.\cite{Eichler09}
Note that $\phi_\Lambda^0$ now also
has an explicit $\Lambda$-dependence which will be chosen such
that $\Gamma_\Lambda^{(1,0)}$ and $\Gamma_\Lambda^{(0,1)}$ 
vanish for all $\Lambda$.\cite{Schuetz2005}
For our purpose it is sufficient to 
approximate the effective action by the following ansatz,
\begin{equation}
  \Gamma_\Lambda [\bar{\phi},\phi]
  \approx\intop_K \bar{\phi}^{}_K \sigma^{}_\Lambda(K) \phi^{}_K 
  + \frac{1}{2} \intop_K 
  \delta\rho^{}_K u^{}_\Lambda(K) \delta\rho^{}_{-K} \, ,
  \label{eq:defgamma}
\end{equation} 
where 
\begin{equation}
  \delta\rho_K=\intop_Q \bar{\phi}_{Q}\phi_{Q+K}
  - \delta_{K,0} \rho_\Lambda^0
\end{equation} 
is the Fourier-transform of the condensate density fluctuation $\rho(X)-\rho_\Lambda^0 =
\bar{\phi}_X\phi_X - \rho_\Lambda^0$. Here, in
analogy with the definition of $K$, we use
$X=({\bm{r},\tau})$, where $\bm{r}$ is a $D$-dimensional real
space coordinate and $\tau$ is the imaginary time.
At the initial UV cutoff scale $\Lambda=\Lambda_0$ Eq.~(\ref{eq:defgamma})   is 
exact and $u_{\Lambda_0}(K)$
is just the bare interaction which enters Eq.~(\ref{eq:InAct})
whereas $\sigma_\Lambda$ vanishes initially,
\begin{equation}
  \sigma_{\Lambda_0}(K)=0 \, .
\end{equation}
 While the
effective action (\ref{eq:defgamma}) describes arbitrarily 
strong density-density 
interactions, processes which 
involve three or
more particles are neglected. Thus, the ansatz is not expected to be accurate 
for dense systems
but it can describe dilute yet strongly interacting ones. Note that the
effective action is completely parameterized in terms of the scalar $\rho_\Lambda^0$
and the two cutoff-dependent
scalar functions
$\sigma_\Lambda(K)$ and $u_\Lambda(K)$.
Eq.~(\ref{eq:defgamma}) represents
a non-local potential approximation of the action and is explicitly 
$U(1)$ invariant for any value of $\Lambda$
even if the condensate density $\rho_\Lambda^0$ is finite. 
In the usual field 
expansion, the $U(1)$ symmetry of the model
leads to Ward identities which relate higher order irreducible 
vertices to lower ones. In
our approach the irreducible vertices are derived from an explicitly 
invariant effective
action which guarantees that all Ward identities are automatically obeyed.
The normal self-energy 
\begin{equation}
\Sigma_\Lambda^N(K)=\Gamma_\Lambda^{(1,1)}(K,K)+\mu
\end{equation}
and the anomalous
self-energy 
\begin{equation}
\Sigma_\Lambda^A(K)=\Gamma_\Lambda^{(2,0)}(-K,K)=\Gamma_\Lambda^{(0,2)}(K,-K)
\end{equation} 
have the form (here and below we shall assume, without loss of generality, 
a real valued condensate wave function $\phi_\Lambda^0$ for notational convenience)
\begin{subequations}
  \begin{eqnarray}
    \Sigma_\Lambda^N(K) & = & \mu + \sigma_\Lambda(K)+ \rho_\Lambda^0 u_\Lambda(K) \, , 
    \label{eq:approxsigman}\\
    \Sigma_\Lambda^A(K) & = & \rho_\Lambda^0 u_\Lambda(K) \, .
    \label{eq:approxsigmaa}
  \end{eqnarray}
\end{subequations} 
Note that
Eqs.~(\ref{eq:approxsigman},\ref{eq:approxsigmaa}) have the
same structure as Eqs.~(\ref{eq:nepomn:SE},\ref{eq:nepomn:NSE}) and
we can already anticipate that $u_\Lambda(K)$ will become
non-analytic for $\Lambda\to 0$ while $\sigma_\Lambda(K)$ will
remain analytic. Moreover, since $\sigma_\Lambda(0)$ vanishes, 
Eqs.~(\ref{eq:approxsigman},\ref{eq:approxsigmaa}) obey
the Hugenholtz-Pines relation (\ref{eq:HPR}) for all $\Lambda$.
Note that Eq.~(\ref{eq:defgamma}) has no terms linear in
the field $\delta \phi$ which for $\Lambda=\Lambda_0$ is achieved
by fixing
the initial
condensate density as
\begin{equation}
\label{eq:ChemPot}
\rho^{0}_{\Lambda_0} = \frac{\mu}{u^{}_{\Lambda^{}_0}}, 
\end{equation}
where 
\begin{equation}
u^{}_{\Lambda^{}_0}=u^{}_{\Lambda^{}_0}(0).
\end{equation} 
%but it will be renormalized on reducing the cutoff.
Eq.~(\ref{eq:defgamma}) also fixes the three- and four-point vertices 
of the theory which have the symmetrized form 
 \begin{eqnarray}
 \Gamma^{(2,1)}_{\Lambda} ( K_1^{\prime} , K_2^{\prime} ; K_1 ) & = &
 \phi^0_{\Lambda} [ u_{\Lambda} ( K_1^{\prime} ) +   u_{\Lambda} ( K_2^{\prime} ) ],
 \label{eq:Gamma21trunc}
 \\
\Gamma^{(1,2)}_{\Lambda} ( K_1^{\prime} ;  K_2 , K_1 ) & = &
 \phi^0_{\Lambda} [ u_{\Lambda} ( K_1 ) +   u_{\Lambda} ( K_2 ) ],
 \hspace{7mm}
 \label{eq:Gamma12trunc}
\\
\Gamma^{(2,2)}_{\Lambda} ( K_1^{\prime} ,K_2^{\prime} ;  K_2 , K_1 ) & = & 
u_{\Lambda} ( K_1^{\prime} - K_1 )  + u_{\Lambda} ( K_2^{\prime} -K_1 ).
\nonumber
 \\
 & &
 \label{eq:Gamma22trunc} 
\end{eqnarray}
These vertices are not independent but are fixed entirely by
$\phi_\Lambda$ and $u_\Lambda(K)$ which can be extracted
from the self-energies. This dependence is again a consequence
of the $U(1)$ symmetry of the model.
% \begin{eqnarray}
% \nonumber
% &\Gamma^{(1,2)}_{\Lambda^{}_0}(K^{}_1;K^{}_2,K^{}_3)=\Gamma^{(2,1)}_{\Lambda^{}_0}(K^{}_1;K^{}_2,K^{}_3)&\\
% \label{eq:Gamma12}
% &  = \delta^{}_{K^{}_1,K^{}_2+K^{}_3}\sqrt{\rho^{}_{0}}\;\left[u^{}_{\Lambda^{}_0}(K^{}_2)+u^{}_{\Lambda^{}_0}(K^{}_3)\right].&
% \end{eqnarray}
% Note that for $\Gamma^{(2,1)}_{\Lambda^{}_0}$, the momentum $K^{}_1$ is associated with the ingoing field, while for $\Gamma^{(1,2)}_{\Lambda^{}_0}$ with the outgoing one. Finally, for the four-legged bare vertex we get
% \begin{eqnarray}
% \nonumber
% &&\Gamma^{(2,2)}_{\Lambda^{}_0}(K^{}_1,K^{}_2;K^{}_3,K^{}_4) = \delta^{}_{K_1+K_2;K_3+K_4}\\
% \label{eq:Gamma22} 
% &&\left[u^{}_{\Lambda^{}_0}(K_1-K_3)+ u^{}_{\Lambda^{}_0}(K_1-K_4)\right].
% \end{eqnarray}
% Here, momenta $K^{}_3$ and $K^{}_4$ are associated with ingoing fields, while momenta $K^{}_1$ and $K^{}_2$ with the outgoing ones. 

For the FRG we also need the cutoff-dependent full propagators
which can again be written in the form of a 
$2\times2-$matrix in field type space \cite{Beliaev58}
\begin{eqnarray}
  {\bm G}_{\Lambda}(K) &=&
  \left( 
    \begin{array}{cc}
      G^{N}_{\Lambda}(K) & G^{A}_{\Lambda}(K)  \\
      G^{A}_{\Lambda}(K)^* & G^{N}_{\Lambda}(-K) 
    \end{array}
  \right) \, ,
  \label{eq:FullPropLambda} 
\end{eqnarray}
where
\begin{subequations}
  \begin{eqnarray}
    G^N_\Lambda(K) &=& 
    \frac{-G^{-1}_{\Lambda,0}(-K)+\Sigma^N_\Lambda(-K)}{{\cal D}_\Lambda(K)}
    \, ,
   \\
    G^A_\Lambda(K) &=& 
    \frac{\Sigma^{A}_\Lambda(K)}{{\cal D}_\Lambda(K)} \, ,
  \end{eqnarray}
\end{subequations}
and the denominator ${\cal D}_\Lambda(K)$ is given by the expression
\begin{eqnarray}
 \hspace{-1.cm}  {\cal D}_\Lambda(K) &=& 
[i\omega+\epsilon_{k}-\mu+\Sigma^{N}_{\Lambda}(-K)]
\nonumber \\
&& \times
  [i\omega-\epsilon_{k}+\mu-\Sigma^{N}_{\Lambda}(K)]
%\nonumber \\ &&
+ |\Sigma^{A}_{\Lambda}(K)|^2 \, .
  \label{eq:Denom2} 
\end{eqnarray}
The exact flow equations for the effective action 
$\Gamma_\Lambda[\bar{\phi},\phi]$ 
has the form \cite{Wetterich93} (here we do not keep track of the flow of the
constant part of $\Gamma_\Lambda[\bar{\phi},\phi]$, 
i.e. the free energy \cite{PKbook})
\begin{equation}
  \partial_\Lambda \Gamma_\Lambda[\bar{\phi},\phi]=\frac{1}{2} \mbox{Tr}
  \Big\{ \partial_\Lambda \bm{G}_{0,\Lambda}^{-1} 
  \big[\bm{\Gamma}^{(2)}_\Lambda- \bm{G}_{0,\Lambda}^{-1} \big]^{-1}
  \Big\} \, ,
  \label{eq:wett}
\end{equation}
where
\begin{eqnarray}
  {\bm \Gamma}^{(2)}_\Lambda[\bar{\phi},\phi] &=& \left(
    \begin{array}{lr}\displaystyle
      \frac{\delta^2\Gamma[\bar{\phi},{\phi}]}{\delta \bar{\phi}\, \delta {\phi}}
      & \displaystyle
      \frac{\delta^2\Gamma[\bar{\phi},{\phi}]}{\delta \bar{\phi} \, \delta \bar{\phi}}
      \\ \displaystyle
      \frac{\delta^2\Gamma[\bar{\phi}, {\phi}]}
      {\delta {\phi}\, \delta {\phi}}
      & \displaystyle
      \frac{\delta^2\Gamma[\bar{\phi}, {\phi}]}{\delta {\phi} \, \delta \bar{ \phi}}
    \end{array}
  \right) \, .
\end{eqnarray}
Eq.~(\ref{eq:wett}), in conjunction with the field expansion 
(\ref{eq:gendef}), 
yields flow equations for the irreducible vertex functions, see e.g.
Refs.~[\onlinecite{Schuetz2005,PKbook}] for details.

\begin{center}
\begin{figure}[t]
\includegraphics[width=8.cm]{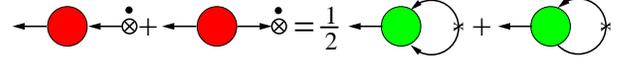}
\caption{(Color online) Diagrammatic representation of the 
  exact FRG flow equation for the condensate density.
  The circles with two external legs on the left-hand side
  denote the normal and
  anomalous self-energy.
  The circles with three external legs on the right-hand side
  denote irreducible three point vertices. 
 %Small black 
  Crossed circles with a dot
  denote the derivative of the order
  parameter with respect to the RG cutoff $\Lambda$,
  and crossed solid arrows represent
  normal or anomalous single scale propagators as defined in
  Eq.~(\ref{eq:defscp}). Arrows pointing out of a vertex represent the fields $\bar{\psi}$,
  while incoming arrows represent fields $\psi$.
}
%    \includegraphics[width=8.cm]{fig1.eps} 
%    \caption{(Color online) Diagrammatic representation of the  
%      exact FRG flow equation for the condensate density. 
%      The shaded circles with two external legs on the left-hand side
%      denote the normal and
%      anomalous self-energy. 
%      The shaded circles with three external legs on the right-hand side 
%      denote irreducible three point vertices. The small black circles with a dot 
%      denote the derivative of the order
%      parameter with respect to the RG cutoff $\Lambda$,
%     and the crossed solid arrows represent
%      normal or anomalous single scale propagators as defined in 
%      Eq.~(\ref{eq:defscp}). Arrows pointing out of a vertex represent the fields $\bar{\psi}$, 
%     while incoming arrows represent $\psi$.
%    }
    \label{fig:Saddle}
  \end{figure}
\end{center}

\begin{center}
  \begin{figure}[t]
    \includegraphics[width=8.5cm]{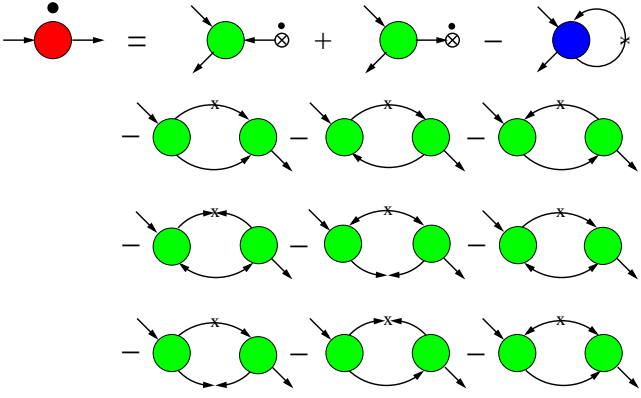}
    \caption{(Color online)
      Diagrammatic representation of the exact FRG flow equation for the normal
      self-energy $\Sigma^N_\Lambda(K)$. The circle with four external
      legs denotes
     the irreducible four-point vertex, while arrows
      represent cutoff-dependent normal or anomalous propagators.
      The dot
      above the two-point vertex on the left-hand side
      represents a derivative with respect to $\Lambda$.
      The other symbols are defined as in Fig.~\ref{fig:Saddle}.}
%    \includegraphics[width=8.5cm]{fig2.eps} 
%    \caption{(Color online) 
%      Diagrammatic representation of the exact FRG flow equation for the normal 
%      self-energy $\Sigma^N_\Lambda(K)$. The shaded circles with our external legs denote 
%     an irreducible four-point vertex, while the arrows
%      represent cutoff-dependent normal or anomalous propagators.
%      The dot
%      above the two-point vertex on the left-hand side 
%      represents a derivative with respect to $\Lambda$.  The
%      other symbols are defined as in Fig.~\ref{fig:Saddle}.}
    \label{fig:Gamma11}
  \end{figure}
\end{center}

\begin{center}
  \begin{figure}[t]
    \includegraphics[width=8.5cm]{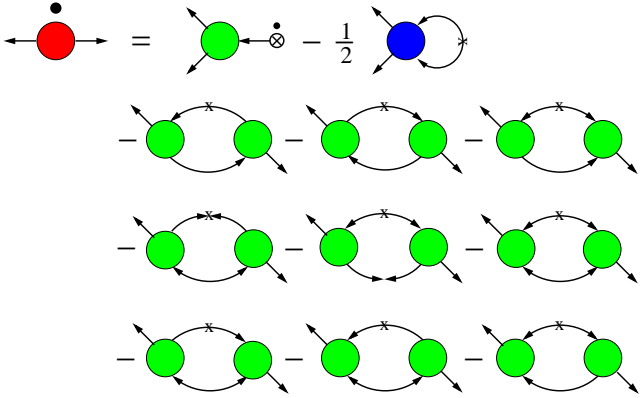} 
    \caption{(Color online) Diagrammatic representation of the exact FRG flow equation for the
      anomalous self-energy $\Sigma_\Lambda^A(K)$.
      All symbols are defined as in Figs.~\ref{fig:Saddle} and \ref{fig:Gamma11}.}
%    \caption{(Color online) Diagrammatic representation of the exact FRG flow equation for the 
%      anomalous self-energy $\Sigma_\Lambda^A(K)$. The symbols have the same meaning
%      as in Figs.~\ref{fig:Saddle} and \ref{fig:Gamma11}.}
    \label{fig:Gamma20}
  \end{figure}
\end{center}

To ensure that the field expansion (\ref{eq:gendef}) is always an expansion
around the true minimum $\phi^0_\Lambda$, we require that $\Gamma_\Lambda^{(1,0)}$
and $\Gamma_\Lambda^{(0,1)}$ vanish for all $\Lambda$. Since for $\Lambda=\Lambda_0$
this is already achieved by the mean field Eq.~(\ref{eq:ChemPot}), we will
only need to enforce 
$\partial_\Lambda \Gamma_\Lambda^{(0,1)}=\partial_\Lambda \Gamma_\Lambda^{(1,0)}=0$.
This yields the flow of the order parameter $\rho^0_\Lambda$,
\begin{eqnarray}
  \label{eq:flowrho}
  \partial_\Lambda \rho^0_\Lambda&=&\frac{1}{u_\Lambda(0)}
  \intop_Q \Big\{ {\dot G}_\Lambda^N(Q)\big[ u_\Lambda(Q)+u_\Lambda(0) \big]
  \nonumber \\
  && \hspace{2cm}
  +\dot{G}_\Lambda^A(Q) u_\Lambda(Q) \big] \Big\} \, ,
\end{eqnarray}
which is shown diagrammatically in Fig.~\ref{fig:Saddle}. The single scale
propagators $\dot{G}_\Lambda^N$ and $\dot{G}_\Lambda^A$ which appear 
in Eq.~(\ref{eq:flowrho}) are defined via the matrix equation
\begin{equation}
  \left(
    \begin{array}{cc}
      \dot{G}_\Lambda^N(K) & \dot{G}_\Lambda^A(K) \\
      \dot{G}_\Lambda^A(K)^* & \dot{G}_\Lambda^N(-K)
    \end{array}
  \right) 
  =-{\bm G}_\Lambda(K) 
  [ \partial_\Lambda{\bm G}_{0,\Lambda}^{-1}(K)] {\bm G}_\Lambda(K) \, ,
\label{eq:defscp}
\end{equation}
where 
\begin{eqnarray}
  {\bm G}_{0,\Lambda}^{-1}(K)&=&\left(
    \begin{array}{cc}
      G_{0,\Lambda}^{-1}(K) & 0\\
      0 & G_{0,\Lambda}^{-1}(-K)
    \end{array}
  \right)\, .
\end{eqnarray}
The exact FRG flow equations
for the normal and anomalous self-energies are shown diagrammatically
in Figs.~\ref{fig:Gamma11}
and \ref{fig:Gamma20}. Using the relations 
Eqs.~(\ref{eq:Gamma21trunc}-\ref{eq:Gamma22trunc}) for the three- and four-point
vertices, the flow equations for the self-energies reduce to
\begin{widetext}
\begin{subequations}
  \begin{eqnarray}
    \nonumber
    \partial_\Lambda\Sigma_\Lambda^N(K) &=& %(\partial_\Lambda \Phi_\Lambda^0)
    \frac{u_\Lambda(K)+u_\Lambda(0)}{u_\Lambda(0)} \intop_Q 
    \big[\dot{G}_\Lambda^N(Q)+ \dot{G}_\Lambda^A(Q)\big] u_\Lambda(Q) +
    \intop_Q \dot{G}_\Lambda^N(Q) \big[u_\Lambda(K)-u_\Lambda(K-Q)\big] 
    \\ && -
    \rho^0_\Lambda \intop_Q \dot{G}_\Lambda^N(Q) \Big\{ G_\Lambda^N(K-Q) 
    \big[u_\Lambda(Q)+u_\Lambda(K-Q)\big]^2 +
    G_\Lambda^N(Q-K)\big[u_\Lambda(K)+u_\Lambda(K-Q)\big]^2 \nonumber 
    \\ && \hspace{.8cm}
    + G_\Lambda^N(Q+K)\big[u_\Lambda(Q)+u_\Lambda(K)\big]^2 
    + 2 G_\Lambda^A(K-Q)\big[u_\Lambda(Q)+u_\Lambda(K-Q)\big]
    \big[u_\Lambda(K)+u_\Lambda(K-Q)\big]
    \Big\} \nonumber
    \\ && 
    -2 \rho^0_\Lambda \intop_Q \dot{G}_\Lambda^A(Q)\Big\{
    G_\Lambda^A(K-Q)\big[u_\Lambda(K)+u_\Lambda(K-Q)\big]
    + G_\Lambda^N(K-Q)\big[u_\Lambda(Q)+u_\Lambda(K-Q)\big]
    \Big\}\nonumber \\ &&
    \hspace{2.6cm}
    \times \, \big[u_\Lambda(K)+u_\Lambda(Q)\big]
    \label{eq:flowscnormal} \, ,
  \end{eqnarray}
  \begin{eqnarray}
    %\\ 
    \nonumber
    \partial_\Lambda \Sigma_\Lambda^A(K)&=&\frac{u_\Lambda(K)}{u_\Lambda(0)}
    \intop_Q \Big\{ \dot{G}_\Lambda^N(Q) \big[u_\Lambda(Q)+u_\Lambda(0)\big]
    +\dot{G}_\Lambda^A(Q) u_\Lambda(Q)
    \Big\}
    -\frac{1}{2}\intop_Q \dot{G}_\Lambda^A(Q)\big[u_\Lambda(K+Q)+u_\Lambda(K-Q)\big]
    \nonumber \\ &&\hspace{-1.8cm}
    -\rho^0_\Lambda \intop_Q \dot{G}_\Lambda^N(Q)\Big\{
    \big[G_\Lambda^N(Q-K)+G_\Lambda^N(K-Q)\big] 
    \big[u_\Lambda(K)+u_\Lambda(K-Q)\big]\big[u_\Lambda(K)+u_\Lambda(Q)\big]
    \nonumber \\ &&   \hspace{.4cm}
    +\Big(G_\Lambda^A(K+Q)\big[u_\Lambda(Q)+u_\Lambda(K+Q)\big]+
    G_\Lambda^A(K+Q)\big[u_\Lambda(Q)+u_\Lambda(K+Q)\big]
    \Big)\big[u_\Lambda(Q)+u_\Lambda(K)\big]
    \Big\}
    \nonumber \\ && \hspace{-1.8cm}
    -\rho^0_\Lambda \intop_Q \dot{G}_\Lambda^A(Q)\Big\{
    G_\Lambda^A(K-Q) 
    \Big(\big[u_\Lambda(Q)+u_\Lambda(K-Q)\big]^2
    +\big[u_\Lambda(K)+u_\Lambda(K-Q)\big]^2\Big)    +G_\Lambda^A(K+Q)\big[u_\Lambda(K)+u_\Lambda(Q)\big]^2
    \nonumber \\ &&
    \hspace{.4cm}
    +\big[ G_\Lambda^N(K-Q)+ G_\Lambda^N(Q-K) \big]
    \big[u_\Lambda(K)+u_\Lambda(K-Q)\big]\big[u_\Lambda(Q)+u_\Lambda(K-Q)\big]
    %\nonumber  \\ &&
%    \hspace{1.4cm}
    \Big\}
    \label{eq:flowscanomalous} \, ,
  \end{eqnarray}
\end{subequations}
\end{widetext} 
which, together with Eqs.~(\ref{eq:approxsigman},\ref{eq:approxsigmaa}), yield
the flow equations for $\sigma_\Lambda(K)$ and 
$u_\Lambda(K)$. We stress that up to this point the only approximation 
we have made
is that we only kept those irreducible correlations which are explicitly stated
in the Eq.~(\ref{eq:defgamma}) for $\Gamma_\Lambda[\bar{\phi},\phi]$, 
i.e. we only consider two-body
density-density correlations. However, these can be arbitrarily strong and
the set 
of Eqs.~(\ref{eq:flowrho},\ref{eq:flowscnormal},\ref{eq:flowscanomalous})
allow to calculate $\rho^0_\Lambda$, $u_\Lambda(K)$, and $\sigma_\Lambda(K)$, 
which completely determine all vertices of Eq.~(\ref{eq:defgamma}), without
any further approximation. In this respect, our approach goes beyond 
the set of equations given in Appendix C of Ref.~[\onlinecite{Dupuis2009b}]
where additional approximations were made.
While of similar complexity,
the equations derived by Dupuis\cite{Dupuis2009b}
do not treat 
the full frequency- and momentum-dependence of the three- and 
four-point vertices, which appear in the flow of
the self-energies, exactly. In some diagrams the dependence of these 
vertices on
the internal frequency and momentum is neglected.

Eqs.~(\ref{eq:flowrho},\ref{eq:flowscnormal},\ref{eq:flowscanomalous}) form
a set of integro-differential equations which in principle must be solved
self-consistently. While similar flow equations were recently
solved exactly for a model of classical crystalline membranes,\cite{Braghin10}
for the present
quantum system the calculation of an exact solution is an extremely 
difficult problem which we will
not attempt to solve. Rather, we will use a
non-self-consistent approximation
which is motivated by the structure of perturbation theory.
Note that second order perturbation theory (including its divergences) 
is recovered
if we replace on the right hand side 
of Eqs.~(\ref{eq:flowrho},\ref{eq:flowscnormal},\ref{eq:flowscanomalous})
the full function $u_\Lambda(K)$ by the constant $u_{\Lambda_0}$ and use
$\sigma_\Lambda=0$.

Before we calculate the full momentum and frequency dependence
of the self-energies using a non-self-consistent scheme, we first
consider a derivative expansion of $\Gamma_\Lambda$ in the next
subsection.

\subsection{Derivative expansion}
\label{sec:derivative}
A derivative expansion of $u_\Lambda(K)$ 
%$\sigma_\Lambda(K)$ 
is problematic, 
since $u_\Lambda(K)$ becomes 
non-analytic for $\Lambda\to 0$. 
%A derivative expansion cannot be applied
%for it and 
We therefore approximate it only by its
constant part,
\begin{equation}
  \label{eq:approxu}
  u_\Lambda(K)\approx u_\Lambda(0)=u_\Lambda \, .
\end{equation}
This leads to a great simplification for the three- and four-point
vertices. %and the flow equations
%The flow equations then follow directly from 
%Eqs.~(\ref{eq:flowrho},\ref{eq:flowscnormal},\ref{eq:flowscanomalous})
%become much simpler.
With the approximation (\ref{eq:approxu}) 
one finds from
Eq.~(\ref{eq:flowrho}) the simplified form for the flow of the
condensate \cite{SHK2008,Eichler09}
\begin{eqnarray}
  \partial_\Lambda \rho^0_\Lambda =
  \intop_{K}\left( 2 \dot G^{N}_\Lambda(K)+\dot G^{A}_\Lambda (K) \right),
  \label{eq:FRGord} 
\end{eqnarray}
and from Eqs.~(\ref{eq:flowscnormal},\ref{eq:flowscanomalous}) and (\ref{eq:FRGord})
the simplified flows for the self-energies,\cite{SHK2009,Eichler09}
\begin{widetext}
  \begin{subequations}
  \begin{eqnarray}
    \partial_\Lambda\Sigma^{N}_\Lambda(K)& = & 2u^{}_\Lambda\intop_{Q}\left\{ 
      \dot{G}^{N}_\Lambda(Q)+\dot{G}_\Lambda^{A}({Q})\right\} 
    \nonumber
    -  4u_\Lambda^2 \rho^0_\Lambda\intop_{Q}\Big\{ \dot{G}_\Lambda^{N}({Q}) 
    \big[G_\Lambda^{N}(Q+K)+G_\Lambda^{N}(Q-K)+G_\Lambda^{N}(-Q+K)       
    \\
    & & + 2 G^{A}_\Lambda(Q-K) \big]
    +2\dot G_\Lambda^{A}(Q) 
    \big[G_\Lambda^{A}(Q+K) + G_\Lambda^{N}(Q+K) \big] \Big\} \, ,       
    \label{eq:NormVertFlow} 
    \\
    \partial_\Lambda\Sigma^{A}_\Lambda(K)& = & 2u_\Lambda\intop_{Q}
    \dot{G}^{N}_\Lambda(Q) 
    -4 u_\Lambda^2 \rho^0_\Lambda \intop_Q \Big\{ 
    \dot{G}^{N}_\Lambda(Q) \big[G_\Lambda^{N}(Q+K)
    +G_\Lambda^{N}(Q-K)+G_\Lambda^{A}(Q+K)
    +G_\Lambda^{A}(Q-K) \big] \nonumber 
    \\
    && +  \dot{G}^{A}_\Lambda(Q) \big[G_\Lambda^{N}(Q+K)+G_\Lambda^{N}(Q-K)
    +3G_\Lambda^{A}(Q+K) \big] \Big\} \, .
    \label{eq:AnomVertFlow}
  \end{eqnarray}
  \end{subequations}
\end{widetext}
We now further employ a derivative expansion of 
$\sigma_\Lambda(K)$, which is analytic also for $\Lambda \to 0$,
where we keep only the
leading order terms in an expansion 
in $\omega$ and $k$. We thus approximate \cite{SHK2009}
\begin{equation}
  \label{eq:SmSigma} 
  \sigma^{}_\Lambda(K)\approx
i\omega\left(1-Y_\Lambda \right)+
\epsilon_k(Z^{-1}_\Lambda-1)+\omega^2V_\Lambda,
\end{equation}
where initially we have $Z_{\Lambda_0}=Y_{\Lambda_0}=1$ and 
$V_{\Lambda_0}=0$.
With this truncation, the normal and anomalous propagators are simply
\begin{subequations}
  \begin{eqnarray}
    G^{N}_{\Lambda}(K) &=& \frac{ Y_\Lambda i\omega   + Z^{-1}_{\Lambda} 
      \epsilon_{k} +\Delta_{\Lambda}  + R_{\Lambda} ( k )+V_\Lambda \omega^2}
      {{\cal D}^{}_\Lambda(K)} \, ,
    \label{eq:GNapprox} \nonumber \\
    \\
    G^{A}_{\Lambda}(K) &=&- 
    \frac{ \Delta_{\Lambda}  }{{\cal D}^{}_\Lambda(K)}
%{ \omega^2 -
%      \Delta_{\Lambda}^2  + [ Z^{-1}_{\Lambda} \epsilon_{\bd{k}} + \Delta_{\Lambda}   -  
%      R_{\Lambda} ( \bd{k} )   ]^2} ,
    \, ,
    \label{eq:GAapprox} 
  \end{eqnarray}
\end{subequations}
where  
%${\cal D}_\Lambda^{\rm de}(K)$ is the derivative approximation of
%${\cal D}_\Lambda(K)$,
\begin{eqnarray}
{\cal D}_\Lambda^{}(K)&=&Y_\Lambda^2 \omega^2+
[V_\Lambda \omega^2 + Z_\Lambda^{-1} \epsilon_k +R_\Lambda(k)] \nonumber \\
&& \hspace{-.4cm} \times
[2 \Delta_\Lambda+V_\Lambda \omega^2+ Z_\Lambda^{-1} \epsilon_k +R_\Lambda(k)]
\, ,
\end{eqnarray}
and we defined 
\begin{equation}
  \Delta_\Lambda=\rho_\Lambda^0 u_\Lambda \, .
\end{equation}
The corresponding  single-scale propagators follow from Eq.~(\ref{eq:defscp}).
% \begin{subequations}
%   \begin{eqnarray}
%     \nonumber
%     \dot G^{N}_\Lambda(K) & = & -\partial_\Lambda 
%     R^{}_\Lambda(k)\left\{[G^{N}_\Lambda(K)]^2+[G^{A}_\Lambda(K)]^2\right\},\\
% \label{eq:dotGNapprox}\\
% \nonumber
% \dot G^{A}_\Lambda(K) & = & -\partial_\Lambda R^{}_\Lambda(k)
% G^{A}_\Lambda(K)\left\{G^{N}_\Lambda(K)+G^{N}_\Lambda(-K)\right\}.\\
% \label{eq:dotGAapprox} 
% \end{eqnarray}
%   \begin{eqnarray}
%     \dot{G}^{N}_{\Lambda} (K) &=& 
%     - [ \partial_{\Lambda}  R_{\Lambda} ( \bd{k} ) ]
%     \nonumber
%     \\
%     & & \hspace{-15mm} \times
%     \frac{ \Delta_{\Lambda}^2 +  [  -i\omega  - Z^{-1}_{\Lambda} 
%       \epsilon_{\bd{k}} -\Delta_{\Lambda}  + R_{\Lambda} ( \bd{k} ) ]^2 
%     }{ \bigl[ \omega^2 -
%       \Delta_{\Lambda}^2  + [ Z^{-1}_{\Lambda} \epsilon_{\bd{k}} 
%       + \Delta_{\Lambda}   -  
%       R_{\Lambda} ( \bd{k} )   ]^2 \bigr]^2 } ,
%     \label{eq:dotGNapprox}
%     \\
%     \dot{G}^{A}_{\Lambda} (K) &=& 
%     - [ \partial_{\Lambda}  R_{\Lambda} ( \bd{k} ) ]
%     \nonumber
%     \\
%     & & \hspace{-15mm} \times
%     \frac{ 2 \Delta_{\Lambda} [ R_{\Lambda} ( \bd{k} )  - Z^{-1}_{\Lambda}  
%       \epsilon_{\bd{k}} -\Delta_\Lambda  ]    
%     }{ \bigl[ \omega^2 -
%       \Delta_{\Lambda}^2  + [ Z^{-1}_{\Lambda} \epsilon_{\bd{k}} 
%       + \Delta_{\Lambda}   -  
%       R_{\Lambda} ( \bd{k} )   ]^2 \bigr]^2 } ,
%     \label{eq:dotGAapprox} 
%   \end{eqnarray}
%\end{subequations}
We now employ the Litim regulator \cite{Litim2001} defined through
\begin{equation}
  \label{eq:litim}
  R_\Lambda(k)=\left(1-\delta_{{k},0}\right)(2mZ^{}_\Lambda)^{-1}
  \left(\Lambda^2-k^2 \right)\Theta\left(\Lambda^2-k^2\right),
\end{equation}
which leads to a simplified form of the flow equations since the
integration over the internal momentum can be performed trivially,
\begin{equation}
  \intop\frac{d^D k}{(2\pi)^D}\partial_\Lambda R_\Lambda (k) 
  {\cal{F}} (k^2)=\kappa^{}_{\Lambda}\frac{\Lambda^{D+1}}{mZ^{}_\Lambda} 
  {\cal{F}} (\Lambda^2),
\label{eq:MomentumInt}
\end{equation}
for any function ${\cal{F}} (k^2)$. Here, we defined 
\begin{equation}
  \kappa_\Lambda=K_D[1-\eta^{z}_\Lambda/(D+2)]/D \, ,
\end{equation}
%with $K_D=2^{1-D}\pi^{-D/2}/\Gamma[D/2]$, 
where 
\begin{equation}
  \eta^{z}_\Lambda=\Lambda \partial_\Lambda \ln Z_\Lambda \, 
\end{equation}
is the scaling dimension of $Z_\Lambda$ and $K_D$ is defined
in Eq.~(\ref{def:Kd}).
From Eqs.~(\ref{eq:FRGord},\ref{eq:AnomVertFlow}) 
we find the flow of $\rho_\Lambda^0$ and $u_\Lambda$
\begin{subequations}
  \begin{eqnarray}
    \partial_\Lambda \rho_\Lambda^0 &=& 
    4  \frac{\Lambda^{D+1}\kappa_\Lambda}{2m Z_\Lambda}\int \frac{d \omega}{2\pi}
    \sum_{n=0}^3\frac{c^{(\rho)}_{n} \omega^{2n}}{{D}_\Lambda^2(i \omega)} 
    \label{eq:rhoFlow} \, , \\
    \partial_\Lambda u_\Lambda &=& 
    4 u_\Lambda^2  \frac{ \Lambda^{D+1}\kappa_\Lambda}{2mZ_\Lambda} \int \frac{d \omega}{2 \pi}\sum_{n=0}^3 
    \frac{c^{(u)}_{n} \omega^{2n}}{{D}_\Lambda^3(i\omega)}, \,\;\;  
\label{eq:uFlow}
\end{eqnarray}
\end{subequations}
where % $D_\Lambda(i\omega)={\cal D}_\Lambda(\Lambda,i \omega)$,
\begin{eqnarray}
\nonumber
&& D_\Lambda(i\omega) = {\cal D}_\Lambda(\Lambda,i \omega)\\
&& = 
Y_\Lambda^2 \omega^2+[\tilde{\epsilon}_\Lambda
  +V_\Lambda \omega^2] 
  [\tilde{\epsilon}_\Lambda+V_\Lambda \omega^2 +2 \Delta_\Lambda], \,\;\, 
\end{eqnarray}
and we used the shorthand 
$\tilde{\epsilon}_\Lambda=\epsilon_\Lambda/Z_\Lambda$.  
The coefficients entering the flow (\ref{eq:rhoFlow}) are
\begin{subequations}
  \begin{eqnarray}
    c^{(\rho)}_0 &=&  
    \tilde{\epsilon}^2_\Lambda
    +\tilde{\epsilon}_\Lambda\Delta_\Lambda+\Delta^2_\Lambda \, ,
    \\
    c^{(\rho)}_1 &=&  
    V_\Lambda(2\tilde{\epsilon}_\Lambda+\Delta_\Lambda)-Y^2_\Lambda \, ,
    \\
    c^{(\rho)}_2 &=& V^2_\Lambda \, ,
  \end{eqnarray}
\end{subequations}
and those entering (\ref{eq:uFlow}) are
\begin{subequations}
\begin{eqnarray}
  \label{eq:cu0}
  c^{(u)}_0 &=&  5\tilde{\epsilon}^3_\Lambda+3\tilde{\epsilon}^2_\Lambda\Delta_\Lambda
  +6\tilde{\epsilon}_\Lambda
  \Delta^2_\Lambda+4\Delta^3_\Lambda \, , \\
  c^{(u)}_1 &=&  3 V_\Lambda \left(5\tilde{\epsilon}^2_\Lambda
    +2\tilde{\epsilon}_\Lambda\Delta_\Lambda+
    2\Delta^2_\Lambda\right) \nonumber \\
  && -Y^2_\Lambda\left(7\Delta_\Lambda+
    11\tilde{\epsilon}_\Lambda\right) \, , \\
  c^{(u)}_2 &=& V_\Lambda \left[3V_\Lambda\left(5\tilde{\epsilon}_\Lambda
      +\Delta_\Lambda\right) -11Y^2_\Lambda\right] \, , \\
  c^{(u)}_3 &=& 5 V^3_\Lambda \, .
\end{eqnarray}
\end{subequations}
To find the flow of the parameters entering $\sigma_\Lambda(K)$, we must expand
the flow of $\Sigma^N_\Lambda(K)$, (\ref{eq:NormVertFlow}), 
to second order in $\omega$ and $k$. This yields
\begin{subequations}
  \begin{eqnarray}
    \hspace{-1.cm}
    \partial^{}_\Lambda Y^{}_\Lambda &=& 
    -8 \rho^0_\Lambda u^{2}_\Lambda Y_\Lambda\frac{\Lambda^{D+1}\kappa_\Lambda}{2mZ^{}_\Lambda}
    \int\frac{d\omega}{2\pi} \sum_{n=0}^{2}\frac{c^{(y)}_{n}\omega^{2n}}
    {{D}_\Lambda^3(i\omega)}  ,
    \label{eq:YFlow} 
    \\ \hspace{-1.cm}
    \partial^{}_\Lambda Z^{}_\Lambda &=& 4 \rho^0_\Lambda u^{2}_\Lambda  
    \frac{\Lambda^{D+1}K_D}{2mD}\int\frac{d\omega}{2\pi}\frac{1}{{D}_\Lambda^2(i\omega)} \, ,
    \label{eq:ZFlow} 
    \\ \hspace{-1.cm}
    \partial^{}_\Lambda V^{}_\Lambda &=& 8 \rho^0_\Lambda u^{2}_\Lambda 
    \frac{\Lambda^{D+1}\kappa_\Lambda}{2mZ^{}_\Lambda}\int\frac{d\omega}{2\pi}\sum_{n=0}^2 
    \frac{c^{(v)}_{n}\omega^{2n}}{{D}_\Lambda^3(i\omega)} \, ,
    \label{eq:VFlow}
  \end{eqnarray}
\end{subequations}
where the coefficients entering the flow of $Y_\Lambda$ are
\begin{subequations}
  \begin{eqnarray}
    c^{(y)}_0 &=&  \tilde{\epsilon}^2_\Lambda-2\tilde{\epsilon}_\Lambda
    \Delta^{}_\Lambda
    -2\Delta^2_\Lambda \, ,\\
    c^{(y)}_1 &=&  Y^2_\Lambda+2(\tilde{\epsilon}_\Lambda-\Delta^{}_\Lambda)
    V^{}_\Lambda \, ,\\
    c^{(y)}_2 &=&  V^2_\Lambda \, , 
  \end{eqnarray}
\end{subequations}
and the coefficients entering the flow of $V_\Lambda$ are 
\begin{subequations}
  \begin{eqnarray}
    c^{(v)}_0 &=& -Y^2_\Lambda(\tilde{\epsilon}_\Lambda+
    \Delta^{}_\Lambda)-\tilde{\epsilon}_\Lambda
    (\tilde{\epsilon}_\Lambda+2\Delta^{}_\Lambda)V^{}_\Lambda \, , \\
    c^{(v)}_1 &=& 2V^{}_\Lambda [Y^2_\Lambda+V^{}_\Lambda(\tilde{\epsilon}_\Lambda
    +\Delta^{}_\Lambda)] \, , \\
    c^{(v)}_2 &=& 3V^3_\Lambda \, .
  \end{eqnarray}
\end{subequations}
Before we analyze the results of the derivative expansion, let us briefly
consider the scaling dimensions of the coupling parameters entering
the theory. If we define the momentum to have scale one, 
$\left[k\right]=1$, and
the dimension of frequency to be equal to the
cutoff dependent dynamical exponent $\left[\omega\right]=z_\Lambda$,  
we find 
\begin{subequations}
  \begin{eqnarray}
    \left[ V_\Lambda \right] &=& 2-2z_\Lambda-\eta_\Lambda^z \, , \\
    \left[ u_\Lambda \right] &=& 4-D-z_\Lambda-2\eta_\Lambda^z \, , \\
    \left[ \rho_\Lambda^0 \right]&=&D-2+z_\Lambda+\eta_\Lambda^z \, , 
  \end{eqnarray}
\end{subequations}
where the scaling dimensions of $Z_\Lambda$ and $V_\Lambda$ are defined by
\begin{subequations}
  \begin{eqnarray}
    \left[ Z_\Lambda \right] &=& \eta_\Lambda^z= \Lambda \partial_\Lambda Z_\Lambda \, , \\
    \left[ Y_\Lambda \right] &=& \eta_\Lambda^y= \Lambda \partial_\Lambda Y_\Lambda \, ,
  \end{eqnarray}
\end{subequations}
which fixes the dynamical exponent,
\begin{equation}
  z_\Lambda=2-\eta_\Lambda^z-\eta_\Lambda^y \, .
\end{equation}
Initially, we have $\eta_{\Lambda_0}^z=\eta_{\Lambda_0}^y=0$ and
$z_{\Lambda_0}=2$ which is the correct dynamical exponent
for non-interacting bosons. It is then useful to give the
bare interaction strength $u_{\Lambda_0}$ 
and the chemical potential 
in dimensionless
form,
\begin{subequations}
  \begin{eqnarray}
    \tilde{u}_0 &=&2m u_{\Lambda_0} \Lambda_0^{D-2} 
    \label{eq:udimless} \, , 
\\
    \tilde{\mu}&= & 2m\mu_{\Lambda_0}\Lambda_0^{-2} \, .
    \label{eq:mudimless}
  \end{eqnarray}
\end{subequations}
Note that at $T=0$, the scaling dimension $\eta_\Lambda^z$ plays
no important role since it vanishes at $\Lambda=0$. However,
at the finite temperature transition of the Bose gas the
critical fluctuations lead to a finite 
$\eta^z=\lim_{\Lambda\to 0}\eta_\Lambda^z$.
This was analyzed in Ref.~[\onlinecite{Eichler09}] where the 
anomalous dimension $\eta^z$ was evaluated at the critical point.
The relevant scaling dimension at $T=0$ is however $\eta^y_\Lambda$ which
controls also the crossover from the $z_\Lambda=2$ regime to the 
$z_\Lambda=1$ regime which is characteristic of the 
Goldstone modes.\cite{Dupuis2007,Wetterich2007}

\begin{figure}[t]
\includegraphics[height=5cm]{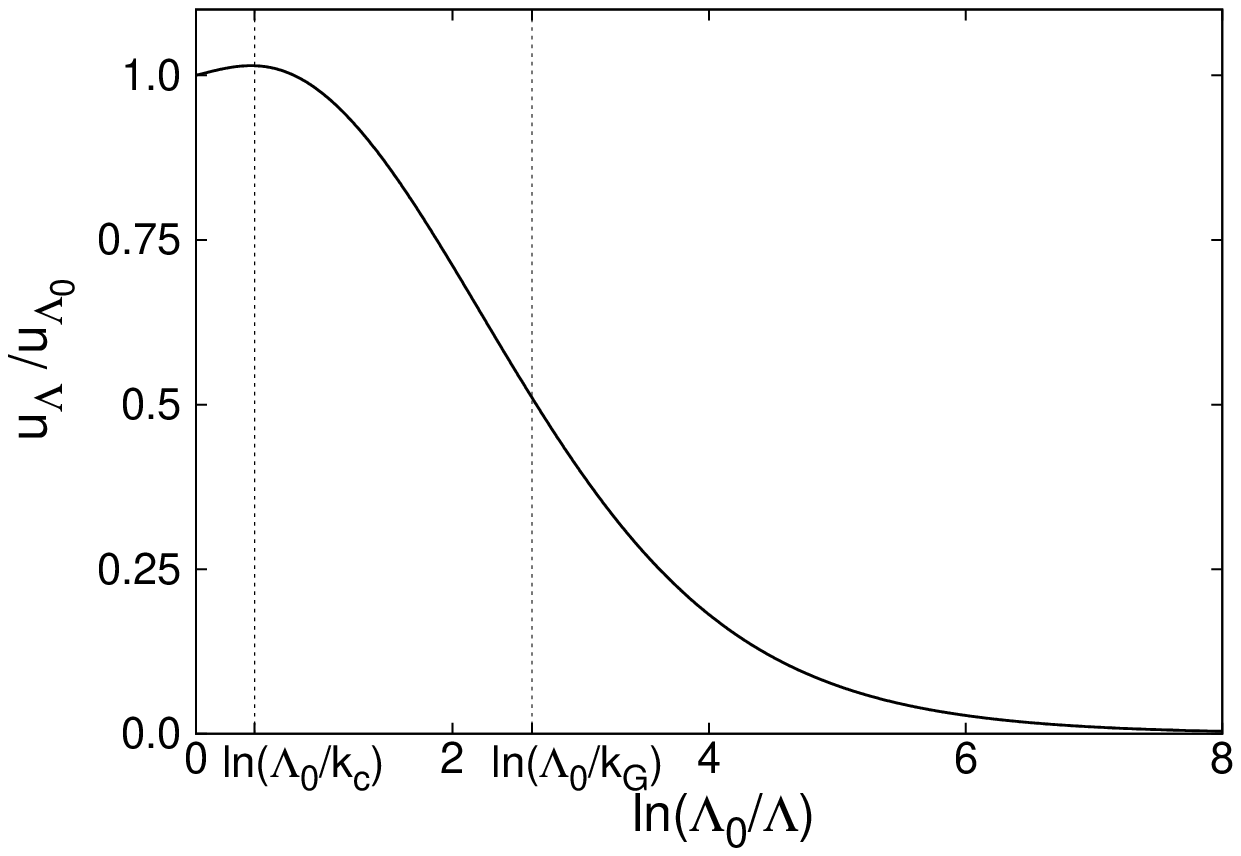}
\includegraphics[height=5cm]{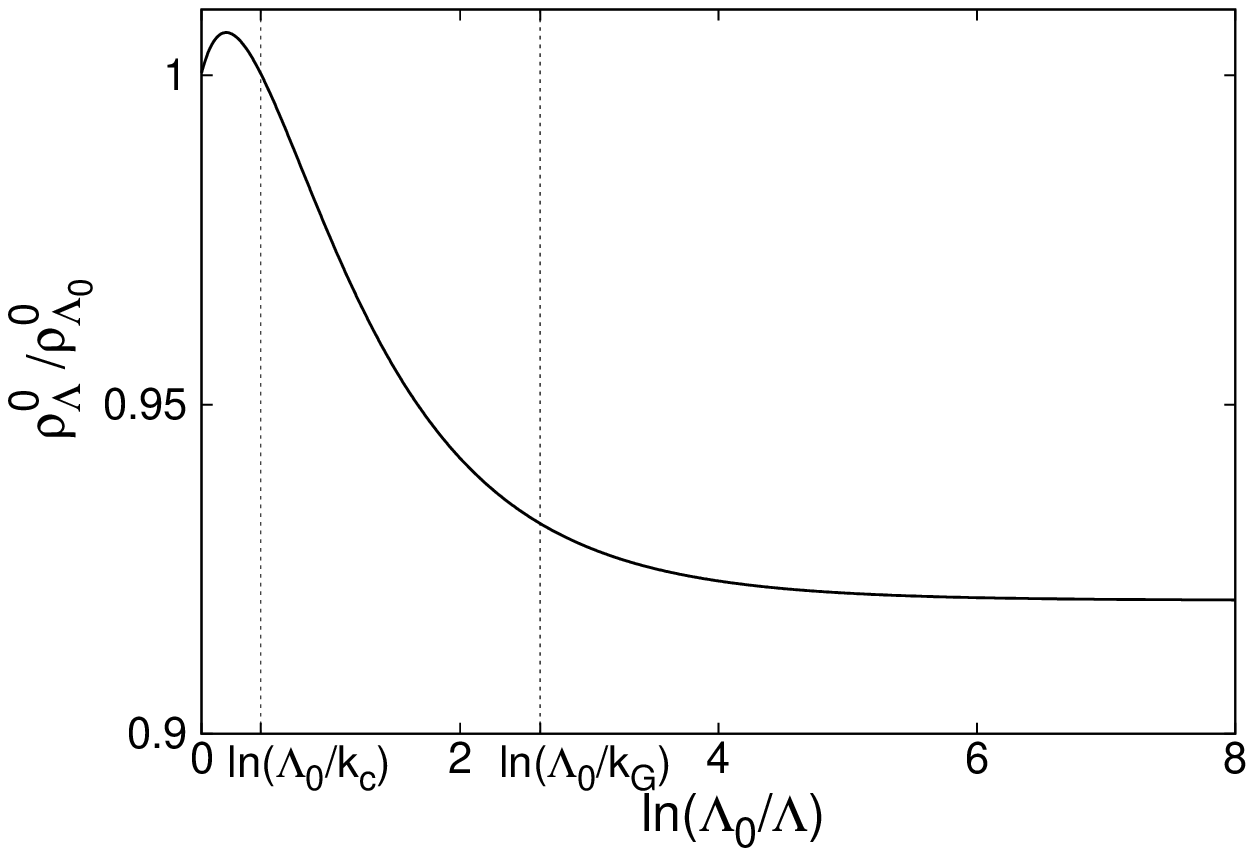}
\caption{Typical RG flows of the interaction parameter
$u^{}_\Lambda$ and the condensate density 
$\rho^{0}_\Lambda$ for $D=2$. The   initial values are 
$\tilde\mu^{}_0=2m\mu\Lambda^{-2}_0=0.4$ and  
$\tilde{u}_0=2m u_{\Lambda_0}\Lambda_0^{D-2}=4$. 
The Ginzburg scale $k_G$ and the crossover scale $k_c = 2 mc$ are indicated 
by vertical lines.}
\label{fig:int}
\end{figure}

\begin{figure}[h]
\includegraphics[height=5cm]{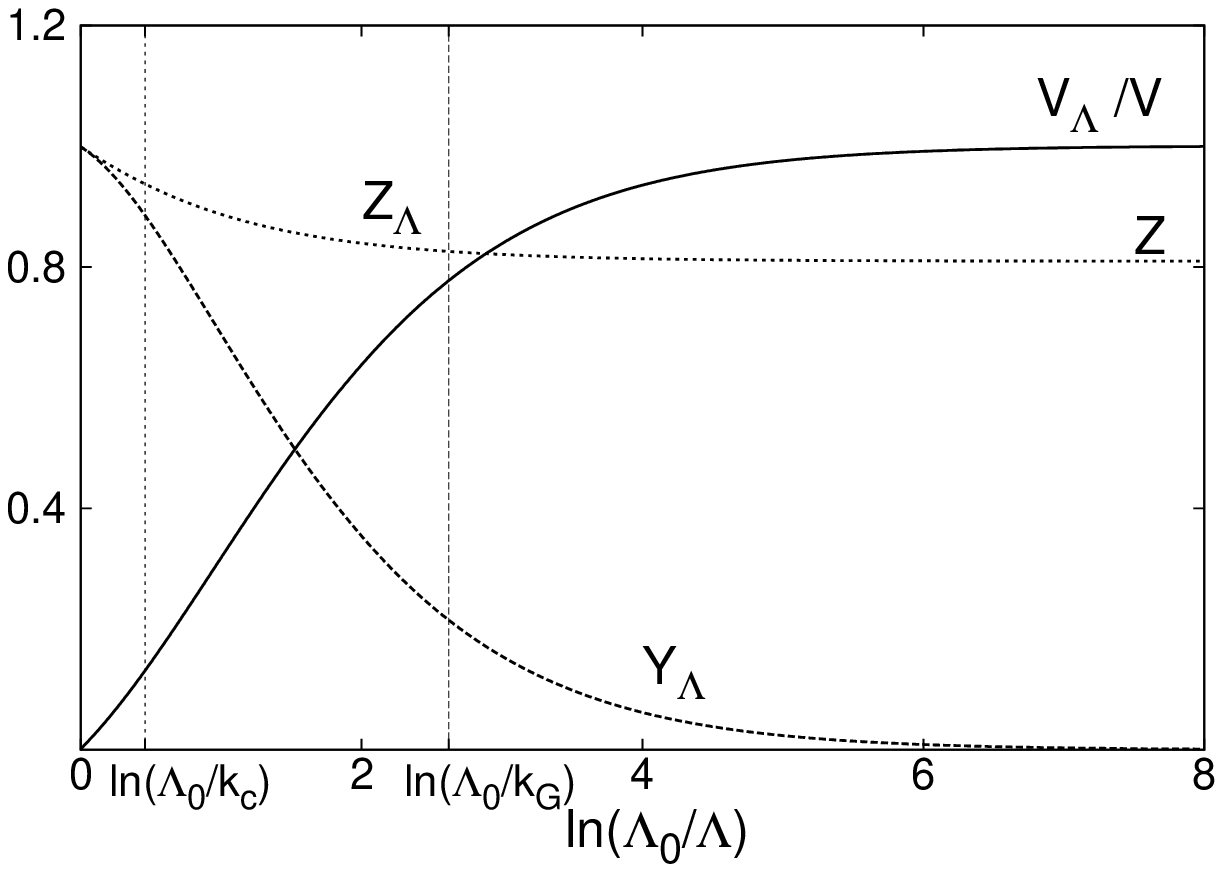}
\includegraphics[height=5cm]{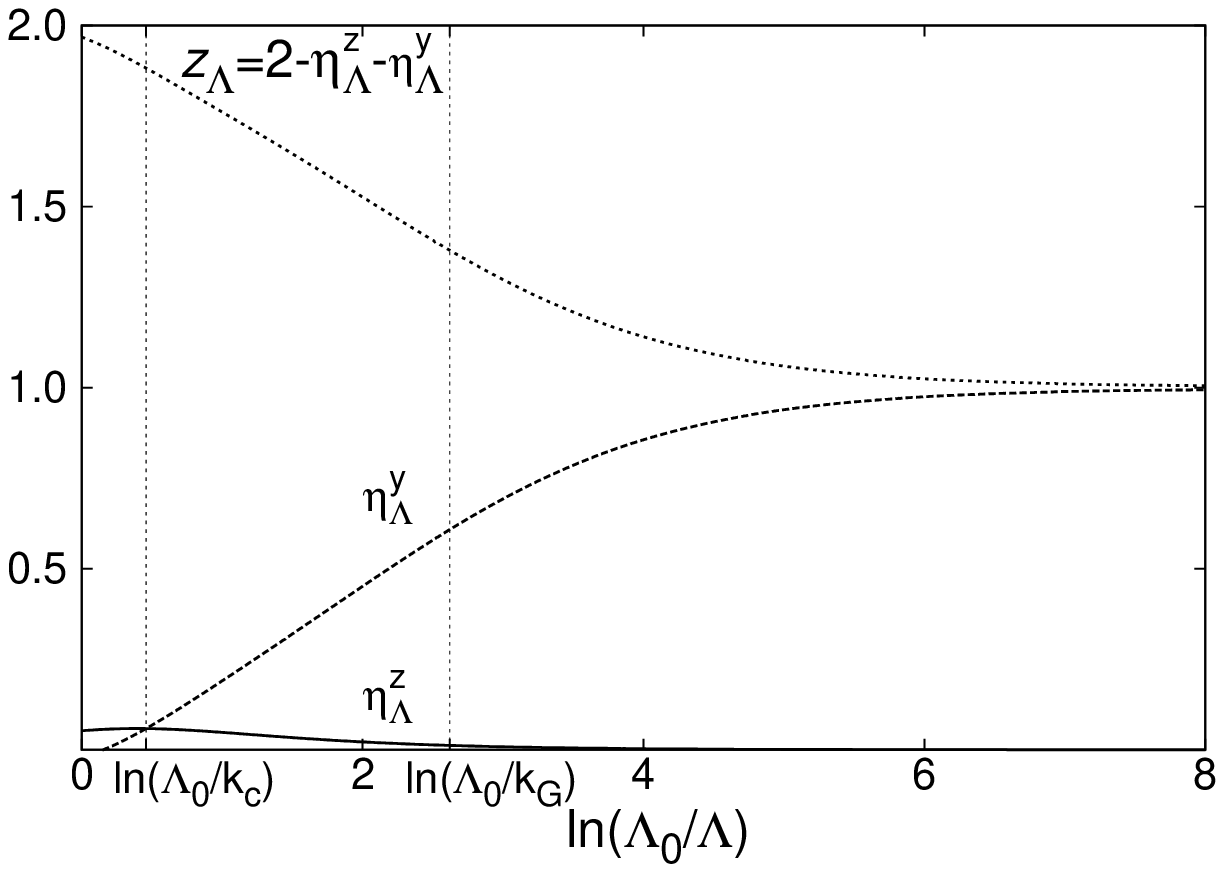}
\caption{Typical RG flows of the coupling parameters 
$Y^{}_\Lambda$, $Z^{}_\Lambda$, $V^{}_\Lambda$ and the dynamical exponent 
$z_\Lambda$ and both anomalous dimensions $\eta^{z}_\Lambda$ and 
$\eta^{y}_\Lambda$ in $D=2$. 
The initial values are the same as in Fig.~\ref{fig:int}.} 
\label{fig:dyn}
\end{figure}

Equations similar to 
Eqs.~(\ref{eq:rhoFlow},
\ref{eq:uFlow}, \ref{eq:YFlow}-\ref{eq:VFlow})
were already discussed in 
Refs.~[\onlinecite{Dupuis2007,Wetterich2007,Dupuis2009,Dupuis2009b,SHK2009,Eichler09}] 
and we briefly summarize the results for $2\leq D \leq 3$.  
At $T=0$ the condensate density $\rho^0_\Lambda$ flows
to some finite limit $\rho^0_{\Lambda\to 0}=\rho^0>0$, 
while
the coupling constant $u_\Lambda$ vanishes for $\Lambda \to 0$ which ensures
$\Sigma^{N}_{\Lambda\to 0}(K=0)=\mu$ and $\Sigma^{A}_{\Lambda\to 0}(K=0)=0$, 
in accordance
with both the Hugenholtz-Pines relation \cite{Hugenholtz1959} 
and the Nepomnyashchy identity.\cite{Nepomnyashchy1975} Furthermore,
$Y^{}_\Lambda$ vanishes for $\Lambda\to 0$, again in accordance with
exact results.\cite{Nepomnyashchy1975} Since $V^{}_\Lambda$ 
and $Z_\Lambda$ are finite for $\Lambda\to 0$, the low energy modes are phonons
with a linear dispersion and 
velocity \cite{Wetterich2007,Dupuis2007,Dupuis2009,Dupuis2009b,SHK2009}
\begin{equation}
\label{eq:GoldVel}
c=\frac{1}{\displaystyle\sqrt{2m V^{} Z^{}}} \, ,
\end{equation}
with $V^{}=V_{\Lambda \to 0}$ and $Z=Z_{\Lambda\to 0}$. 
In Fig.~\ref{fig:int} and Fig.~\ref{fig:dyn} we show for $D=2$ 
the flows for the coupling parameters and for  
the dynamical exponent $z^{}_\Lambda = 2-\eta^{y}_\Lambda-\eta^{z}_\Lambda$ 
which illustrates the crossover from the free particle regime with 
$z_\Lambda=2$ to the Goldstone regime, where $z_\Lambda=1$. 
As pointed out by Dupuis in Refs.~[\onlinecite{Dupuis2009,Dupuis2009b}],
the crossover of $z_\Lambda$ is governed by the generalized Ginzburg scale 
$k_G$, which is indicated in Fig.~\ref{fig:int} 
and Fig.~\ref{fig:dyn} and is extracted from the flow equations 
in Appendix~\ref{sec:GizbScale}. 
Also indicated is the crossover scale $k_c=2mc$.
Evidently, $k_G$ is the relevant scale  controlling the flow of the
dynamical exponent $z_\Lambda$.\cite{Dupuis2009,Dupuis2009b}
On the other hand, the scale which determines the crossover from a
linear to a quadratic behavior in the full
{\em quasiparticle dispersion} is not $k_G$ but the
crossover scale
$k_c$, as is expected from standard perturbative approaches and
as we will show in the next section when we calculate
the
full frequency and momentum dependence of the self-energies.
It is perhaps surprising that the crossover in the momentum dependence
of the quasiparticle dispersion happens
at a completely different scale as the crossover
in the $\Lambda$-dependence of the parameters
entering the derivative expansion. In the weak-coupling regime
this is however easily understood, since a quasiparticle with momentum $k$
in the range
$k_G \ll k \ll k_c$ 
feels the largely unrenormalized bare value $u_{\Lambda_0}$ of the interaction
potential which is sufficient to render its dispersion linear, 
whereas the flow of the parameters entering the derivative expansion
are sensitive only to the non-perturbative effects associated with
the emergence of non-analyticity at scales of the order of the Ginzburg scale $k_G$. 

The lowest order derivative expansion is thus incapable of describing the
quasiparticle dispersion beyond the asymptotic limit. Strictly
speaking, even this limit is not correctly reproduced since
the derivative expansion yields 
the unphysical result $\lim_{\Lambda\to 0}\Sigma_\Lambda^A(K)\approx 
\lim_{\Lambda\to 0}\rho^0_\Lambda u_\Lambda =0$ for all $K$,
and as a consequence
$\lim_{\Lambda \to 0} G_\Lambda^A(K) \to 0$. This 
violates an exact result by  Gavoret and Nozi\`{e}res \cite{Gavoret1964}
which is given in Eq.~(\ref{eq:gavoret}).
Nonetheless, it is possible to extract the correct asymptotic form
of the propagators if one cuts off the flow at a finite $\Lambda$ and uses
$\Lambda\approx \sqrt{\omega^2+c^2 k^2}$, see 
Refs.~[\onlinecite{Dupuis2007,Dupuis2009,Dupuis2009b}].

\section{Excitation energy and damping of quasiparticles}
\label{sec:Spectral}

We now turn to the calculation of the complete frequency and momentum
dependence of the self-energies from the FRG equations. As pointed out 
in Sec.~\ref{sec:FRG}, a completely self-consistent solution of
the flow equations (\ref{eq:flowrho}, \ref{eq:flowscnormal},
\ref{eq:flowscanomalous}) is numerically extremely difficult and
we therefore use an approximation in which we treat only the
coupling parameters of the derivative expansion of Sec.~\ref{sec:derivative} 
self-consistently, while all higher
powers of frequency and momentum are determined in a non-self-consistent
manner. A similar truncation has previously been tested for the
symmetry-broken phase of a $\phi^4$ theory.\cite{SHK2008}
A non-self-consistent solution can be quite simply  achieved, 
starting from the approximate
flow equations given in Eqs.~(\ref{eq:NormVertFlow}, \ref{eq:AnomVertFlow}).
On the right hand side of these equations we use the approximate form of the
propagators given in Eqs.~(\ref{eq:GNapprox}, \ref{eq:GAapprox}) 
which are based on the derivative expansion, see Eq.~(\ref{eq:approxu})
and Eq.~(\ref{eq:SmSigma}). The same approximation is employed for
the single scale propagators. 
%for which we use the expressions given
%in Eqs.~(\ref{eq:dotGNapprox}, \ref{eq:dotGAapprox}). 
Note that Eqs.~(\ref{eq:GNapprox}, \ref{eq:GAapprox}) have a sensible
large $K$ behavior which for any finite $\Lambda$ and all $k>\Lambda$
are similar to the
standard first order result of the Bogoliubov approximation but
with a renormalized interaction parameter $u_\Lambda$, a renormalized
condensate density $\rho^0_\Lambda$, a 
mass renormalization expressed by $Z_\Lambda$ and further a renormalized
frequency dependence as expressed through $Y_\Lambda$ and $V_\Lambda$.

Using again the Litim regulator~(\ref{eq:litim}), 
the flows of the coupling parameters $u_\Lambda$, $\rho_\Lambda$,
$V_\Lambda$, $Z_\Lambda$ and $Y_\Lambda$ can be simply obtained
from the solutions of Eqs.~(\ref{eq:rhoFlow},
\ref{eq:uFlow}, \ref{eq:YFlow}-\ref{eq:VFlow}).
With these approximations, all parameters entering the right hand side
of Eqs.~(\ref{eq:NormVertFlow}, \ref{eq:AnomVertFlow}) are 
completely determined. For any $K$, we can thus simply perform
the integration over $\Lambda$. With appropriate
boundary conditions this yields both $\Sigma^N(K)$ and $\Sigma^A(K)$.
Because of the structural similarity of both the diagrams and
the propagators in this approximation to the Beliaev theory, we
expect to reproduce qualitatively Beliaev's 
perturbative results for both the spectrum and the damping,
bar their divergences.
However, since already the derivative expansion contains the 
key information about the non-analytic structure of the self-energies,
our approach also captures these, as discussed in more detail
in Appendix~\ref{sec:AsymGreen}.

All information about the quasiparticle properties is encoded
in the single-particle spectral density which is related to the imaginary
part of the real-frequency normal Green's function,
%\cite{Pitaevskii,Abrikosov,Fetter}
\begin{equation}
  \label{eq:SpecDens} 
  A(k,\omega) = -2 {\rm Im} G^{N}(k,i\omega\rightarrow\omega + i0).
\end{equation}
Its calculation requires 
an analytical continuation 
to real frequencies. 
Here we apply the standard Pad\'{e} approximant technique \cite{Vidberg}
which has the advantage that it can be easily implemented numerically.
Although it has been originally proposed for dealing 
with systems at finite temperatures, it can also be used for 
zero-temperature calculation if the input functions are sufficiently 
well resolved.
If it does converge, the technique is furthermore quite accurate. Applied
to our problem it proved to be very robust and the results for the spectral
density quickly converged for momenta which are not too small. 
For the actual data used to calculate
the spectral density we used 450 Matsubara frequency
data points for each momentum $k$. Even though the damping is quite
small, this allowed to accurately extract the width of the resonance
and the damping.

% This task requires first determining the full 
% momentum dependence of self-energies. Generally, a momentum dependent 
% physical vertex $\Gamma^{}_0(K)$ can be found from the corresponding flow equation by integrating out the IR-cutoff $\Lambda$
% \begin{equation}
% \label{eq:FRGEPT} 
% \Gamma^{}_0(K) = \Gamma^{}_{\Lambda^{}_0}(K)-\intop_{0}^{\Lambda_0}d\Lambda~\dot\gamma^{}_\Lambda(K),
% \end{equation}
% where $\dot\gamma^{}_\Lambda(K)$ denotes a right hand side of the corresponding flow equation and $\Gamma^{}_{\Lambda^{}_0}(K)$ denotes the vertex at the beginning of the flow. Usually, a corresponding mean-field value can be used as initial condition for the respective vertex. An attempt to solve this problem analytically fails because of mathematical difficulties. Therefore we decided to solve the problem in a purely numerical way.

\begin{figure}[t]
\center
\includegraphics[height=5cm]{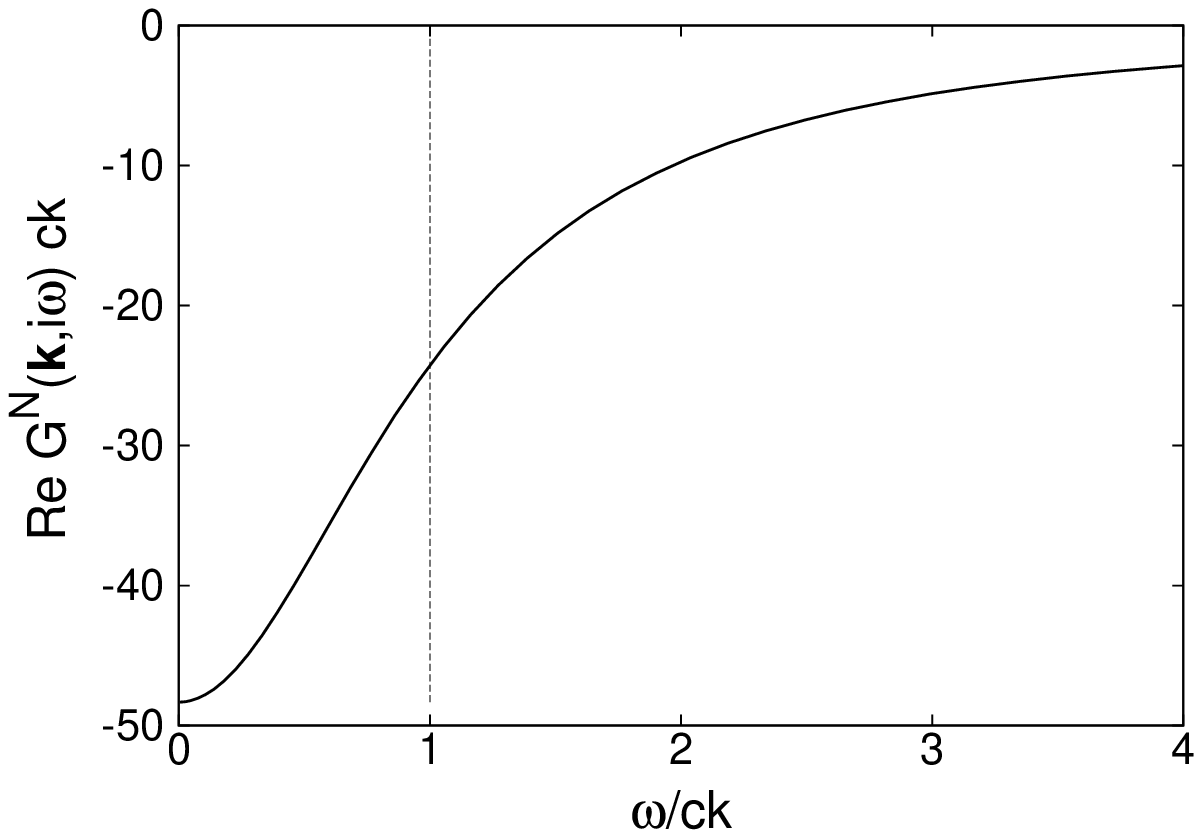} 
\includegraphics[height=5cm]{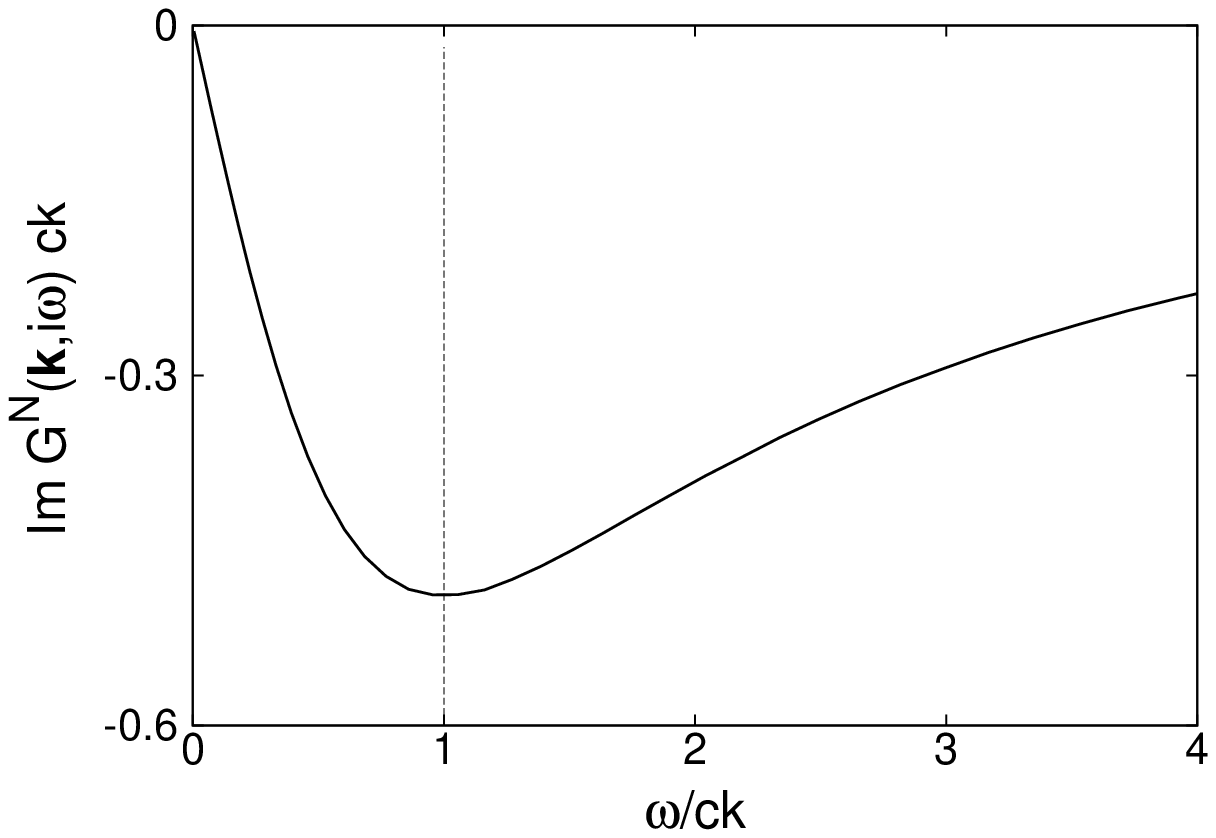} 
\caption{Real and imaginary parts of the normal Matsubara Green's function
as a function of $\omega$, 
calculated within the non-self-consistent FRG approach. The  parameters  
are $\tilde{u}^{}_0=0.8$ and $\tilde{\mu}^{}_0=0.08$, which 
yields $k^{}_c/k^{}_G\approx45$.
%, where $k^{}_c=2mc$ and $k^{}_G$ is defined 
%in Eq.~(\ref{eq:int:GinzScDef}). 
The Green's function is calculated 
for $k = 0.01 k^{}_c\approx 0.45 k^{}_G$. The dashed vertical line shows the 
position of the minimum of the imaginary part, which coincides with 
the predicted resonance at $\omega = c k$.}
\label{fig:MatsGreenFunc}
\end{figure}

The first task is to calculate the Matsubara normal Green's function with 
satisfactory resolution.
A typical result for the normal Matsubara Green's function calculated within 
the non-self-consistent FRG scheme just described is shown in 
Fig.~\ref{fig:MatsGreenFunc}. The initial conditions 
for the condensate and interaction were chosen in such a way that the ratio 
$k^{}_c/k^{}_G$ remains sufficiently large 
(see legend below Fig.~\ref{fig:IntComp}). We further kept the 
ratio $c/c^{}_0$ close to unity to ensure that we are still in the
weakly interacting regime.
For very small momenta (we analyzed momenta down to $k=0.045k^{}_G$ 
for $k^{}_c/k^{}_G\approx 45$), the real part of the normal Matsubara Green's 
function indicates a resonance at $\omega=c k$; however for such
small $k$ the analytic continuation no longer converged.
%excellent accordance with the approximate formula Eq.~(\ref{eq:PropExact}), 
%whose only characteristic energy scale corresponds to the resonance 
%frequency in the real frequency space $\omega=c k$. 
%Moreover, Eq.~(\ref{eq:PropExact}) provides an excellent approximation 
%also for dense systems, for which $c^{}/c^{}_0\ll1$. 
We never found any trace of the Ginzburg scale 
$k^{}_G$ in the properties of the normal Green's function, which,
at least in the weakly interacting regime, 
is perhaps expected since 
in the asymptotic regime the non-analytic terms are known to 
cancel out in the normal Green's functions, see the discussion in
Appendix~\ref{sec:AsymGreen}.
%which is known to play a 
%crucial role for the transversal fluctuations \cite{Dupuis2009}.

In order to check convergence of  the Pad\'{e} approach
we calculated Matsubara Green's functions for up to 450 
non-equidistant frequencies 
in the range $0<\omega< 160c k$ for each $k$ and investigated the 
stability of the results on increasing the number of Pad\'{e} polynomials.
For example, for $k_G\approx 0.2 k_c $ and $k\geqslant0.1k^{}_c$ 
we obtain excellent convergence
and a spectral density which obeys the correct normalization
\begin{equation}
  \intop_{-\infty}^{\infty} \frac{d\omega}{2 \pi}A(k,\omega) =1 
\end{equation}
to high accuracy.
The convergence properties of the Pad\'{e} approximation become continuously
worse
on lowering $k$ and we could not get a satisfactory analytic
continuation in the non-perturbative regime $k\ll k_G$. However, 
this may be simply 
due to the narrowing of the peak width for small momenta. For all momenta 
where we do get a stable analytical continuation we find a damping which
agrees qualitatively with the expected Beliaev damping, even for larger interaction
strengths.
%the number of frequencies between $0<2kc$ is $36$ 
%for each $k$. 

%becomes smaller. An excellent convergence is observable for $k\geqslant0.1k^{}_c$. 
%For these momenta we always reproduce Beliaev damping of quasi-particles 
%Eqs.~(\ref{eq:BelDampGen}) and (\ref{eq:BelDamp2D}) if initial conditions are 
%chosen corresponding to the dilute regime and $k^3-$dependence of the damping 
%in general, regardless the strength of the initial interaction. 

\begin{figure}[t]
\center
\includegraphics[height=5cm]{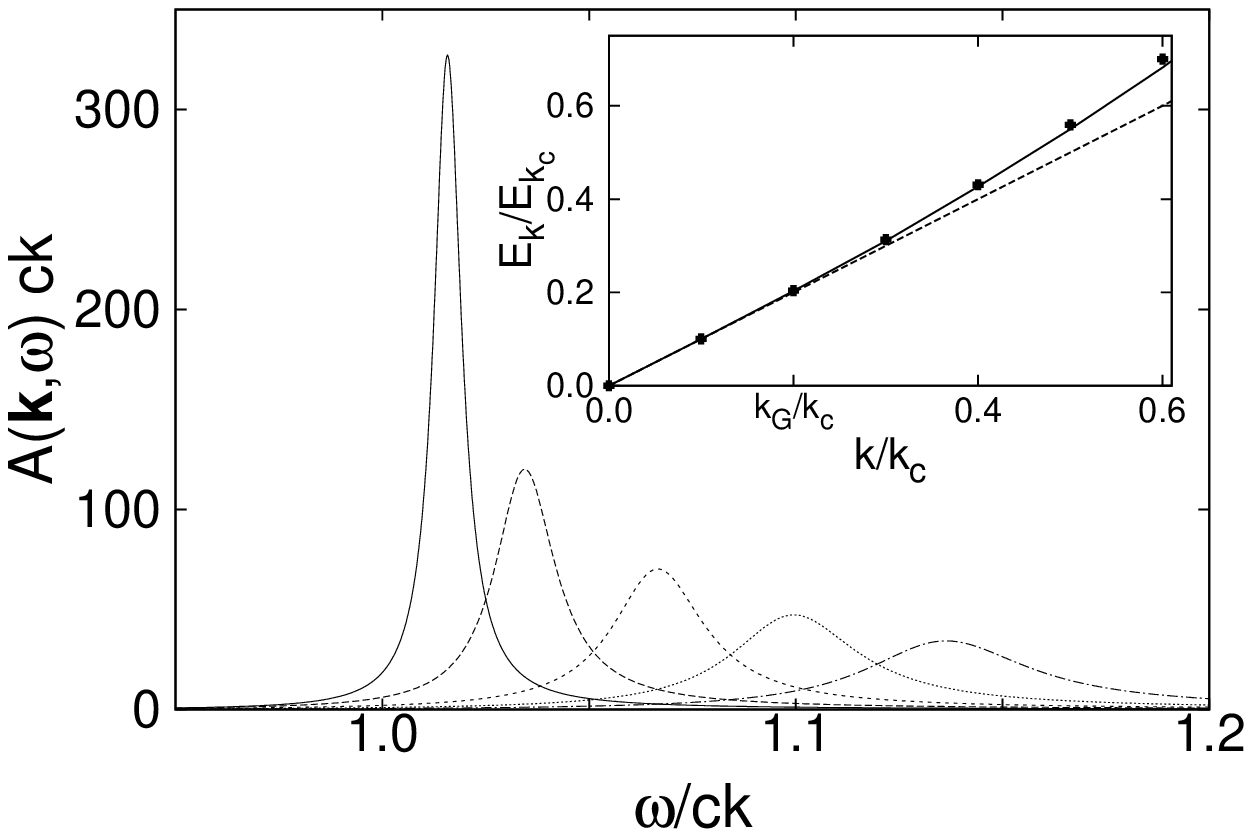}
%\hspace{1.cm}
\includegraphics[height=5cm]{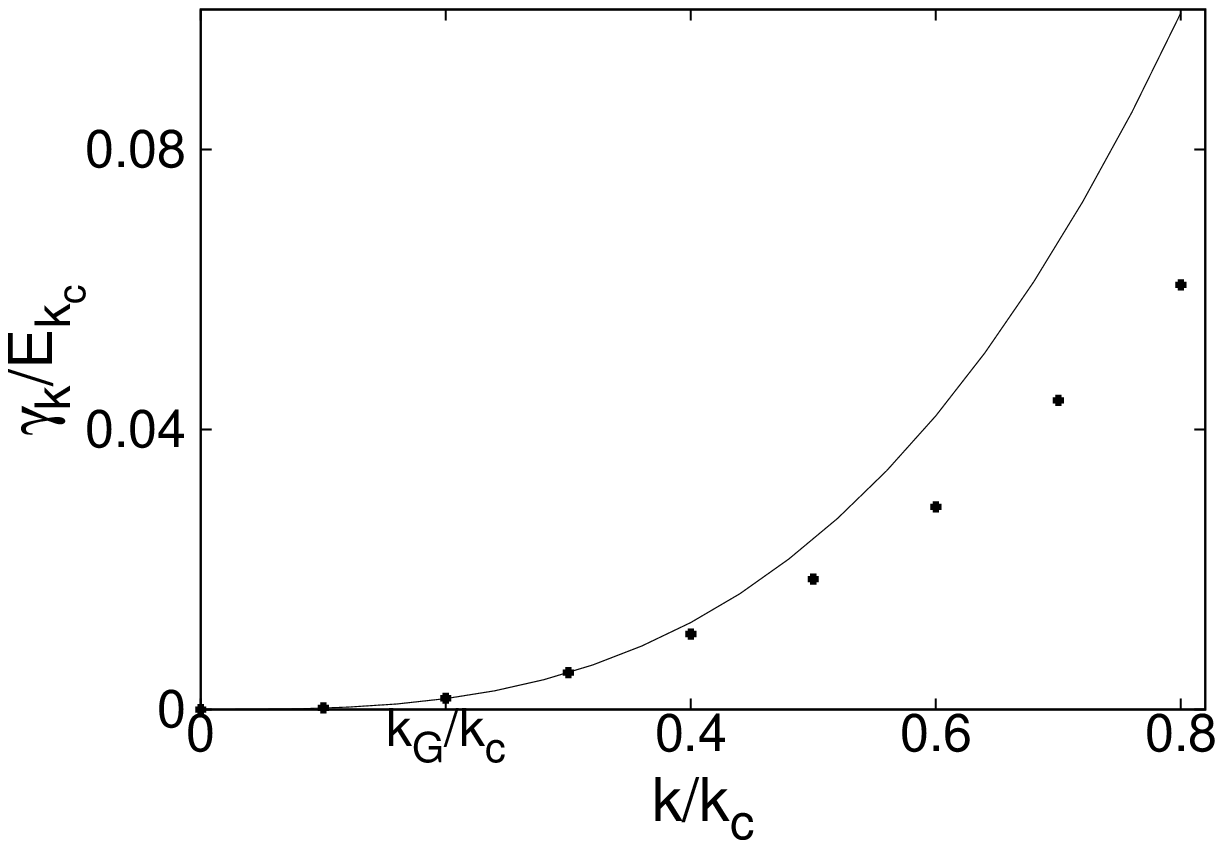}
\caption{FRG results for the single-particle spectral density $A({k},\omega)$ as a function 
of the real frequency  $\omega$ for different values of $k$ (top figure) and 
quasiparticle damping $\gamma_k$ (bottom figure). The shown results are for 
$\tilde\mu_0 =0.15$ and $\tilde{u}_0=15$. In this case  $k_G/k_c \approx 0.2$ which is indicated 
by the labels in the figures. 
The inset in the top figure shows the quasiparticle dispersion $E_k$ 
which deviates at large $k$ from linearity but is well described by a Bogoliubov 
type expression $E_k~=~\sqrt{\epsilon_k^2+ c^2 k^2}$ with renormalized 
velocity $c=(2m V Z)^{-1/2}$ (black dots). The peaks of the spectral 
function correspond to (from left to right)  $k/k_c=0.2,$ $0.3,\, 0.4, 0.5$ and 
$0.6$, where $k_c=2mc$. In the lower plot, black dots are extracted from the spectral density, 
while the solid line fits them as $\gamma_k\approx 0.194k^3/2mk_{c}$.
%
%Single-particle spectral density $A({k},\omega)$ as a function 
%of the real frequency  $\omega$ for different values of $k$.
%, calculated for $ \tilde \mu_0=0.15$ and $\tilde u_0=15$ 
%We show results for $\tilde\mu_0 =0.15$ and $\tilde{u}_0=15$ where
%$k_G/k_c \approx 0.2$, as indicated by the label in the inset and in the bottom figures,
%and the FRG result for the quasiparticle damping $\gamma_k$ (bottom figure).
%The inset in the top figure shows the quasiparticle dispersion $E_k$ 
%which deviates at large $k$ from linearity but is well described by a Bogoliubov 
%type expression $E_k~=~\sqrt{\epsilon_k^2+ c^2 k^2}$ with renormalized 
%velocity $c=(2m V Z)^{-1/2}$ (black dots). The peaks of the spectral 
%function correspond to (from left to right)  $k/k_c=0.2,$ $0.3,\, 0.4, 0.5$ and 
%$0.6$, where $k_c=2mc$.
%In the lower plot, black dots are extracted from the spectral density, 
%while the solid line fits them as $\gamma_k\approx 0.194k^3/2mk_{c}$.
} 
\label{fig:damp}
\end{figure}

A typical shape of the spectral density function 
obtained from the Pad\'{e} approximation is shown in 
Fig.~\ref{fig:damp} for positive $\omega$. One clearly observes a finite 
peak broadening which can be ascribed to Beliaev damping.\cite{Beliaev58} 
The extracted spectrum of elementary excitations is always well fitted by a 
Bogoliubov-like expression $E_k=\sqrt{\epsilon_k^2+ c^2 k^2}$ and the 
damping of quasiparticles always reveals a $k^3$ behavior for $k\to 0$, 
in accordance with the predictions of the perturbative analysis, 
compare Eq.~(\ref{eq:BelDampGen}). However, the prefactor $\alpha_0$ 
introduced in  Eq.~(\ref{eq:BelDampGen}) should be replaced by a function 
$\alpha(\tilde \mu_0,\tilde u_0)$ of the relevant dimensionless parameters 
$\tilde \mu_0$ and $\tilde u_0$ of the model (\ref{eq:InAct}),
\begin{equation}
  \gamma_k^{(D=2)} \approx \alpha(\tilde \mu_0,\tilde u_0) k^3 \, .
\end{equation}
For small 
values of both dimensionless parameters $\tilde{\mu}_0$ and $\tilde{u}_0$
our results for the damping are very
close to the perturbative 
result Eq.~(\ref{eq:BelDamp2D}). For instance, for $\tilde \mu_0=0.008$
 and $\tilde u_0=0.8$ we obtain $c/c_0\approx 1.0054$ and 
$\alpha(\tilde \mu_0,\tilde u_0)/\alpha_0\approx0.967$. 
As both parameters increase (at fixed UV-cutoff $\Lambda_0$), 
we find a stronger renormalization of both the velocity $c$  
of the Goldstone mode and of 
$\alpha(\tilde \mu_0,\tilde u_0)$.
For instance, for $\tilde \mu_0=0.4$ and $\tilde u_0=4$ we 
obtain $c/c_0\approx 1.01$ and $\alpha(\tilde \mu_0,\tilde u_0)/\alpha_0\approx0.915$, 
while the results for $\tilde \mu_0=0.15$ and $\tilde u_0=15$ are  $c/c_0\approx0.669$ 
and $\alpha(\tilde \mu_0,\tilde u_0)/\alpha_0\approx 0.526$. 
The first examples correspond to the weakly interacting regime, where
many-body renormalizations are expected not to play an important 
role, while the last choice of the initial conditions corresponds to a
more strongly interacting Bose gas where deviations from standard
perturbation theory are already noticeable.

\section{Summary and conclusion}
\label{sec:summary}
In conclusion, we have shown how the FRG formalism can be applied to  
calculate the single-particle spectral density
of interacting bosons in dimensions $D\leq 3$. 
We here concentrated on the zero-temperature limit, 
but it is straightforward to extend the analysis also to 
finite temperatures and even to the  critical temperature.\cite{Eichler09}
The FRG seems to be the only presently available analytical
technique which is capable of bridging the 
gap from the non-perturbative regime at very small frequencies and momenta,
where the self-energies become non-analytic, and the intermediate to large
momentum regime which is well described by the standard Beliaev 
theory.\cite{Beliaev58}
In addition to satisfying all symmetry constraints, our approach 
respects the Nepomnyashchy relation
for  the anomalous self-energy~\cite{Nepomnyashchy1975} 
and yields the correct
asymptotic structure of the Green's function as first derived by
Gavoret and Nozi\`{e}res.\cite{Gavoret1964}
Moreover, our approach not only reproduces the 
non-analytic behavior of the self-energies
for scales smaller than the generalized Ginzburg scale $k_G$, 
but it also yields
the standard results for the $T$-matrix renormalization 
in $D=2$ and $D=3$ at momentum scales larger than $k_c = 2 mc$, see
Appendix~B. 
Our approach therefore provides a single unified framework to 
describe the crossover from the long-wavelength Goldstone mode regime 
to the quasi-free
boson regime where the energy dispersion is quadratic.
The correct momentum
dependence of the Beliaev damping, including its prefactor, can be extracted 
from the numerical analysis of the our truncated FRG flow equations. 
While we presented numerical results 
for the single-particle spectral density only for $D=2$, it is straightforward
to apply the technique also in three dimensions.

The explicit results for the spectral function presented in
this work were obtained within a non-self-consistent solution of the
truncated FRG flow equations 
(\ref{eq:NormVertFlow},\ref{eq:AnomVertFlow}) 
for the self-energies.
Although the derivative expansion has served as a guide for the derivation
of Eqs.~(\ref{eq:NormVertFlow},\ref{eq:AnomVertFlow}), our truncation scheme
transcends the derivative expansion because
on the right-hand side of our truncated FRG flow equations all
powers of momenta and frequencies are retained.
See Ref.~[\onlinecite{SHK2008}] for a similar truncation strategy
in the symmetry broken phase of  classical $\phi^4$-theory.

Clearly, our truncation strategy will eventually fail
at strong coupling. However, in principle the FRG approach, which is 
a non-perturbative
RG technique, is well positioned also for the strong coupling regime. 
In fact, our truncated FRG flow equations
(\ref{eq:flowrho},\ref{eq:flowscnormal},\ref{eq:flowscanomalous}) 
takes a two-body density-density interaction of arbitrary strength into account, 
but neglects all higher order (irreducible)  correlations involving
three or more particles. A solution of these
flow equations, or an improved approximative solution, is nonetheless
expected to give valuable insights to strongly coupled bosons.

\section*{ACKNOWLEDGEMENTS}

We thank H. O. Jeschke for introducing us to the Pad\'{e} approximant technique 
and providing us with his routines, and A. L. Chernyshev and N. Dupuis for discussions. 
We acknowledge support by a DAAD/CAPES PROBRAL grant and AS and PK acknowledge support by the SFB/TRR49.
\newline

\appendix
\section{Elements of Nepomnyashchy theory}
\label{app:Nepo}
Gavoret and Nozi\`{e}res showed that both normal 
and anomalous propagators have the same low-energy asymptotic behavior 
with merely a different sign, \cite{Gavoret1964}
\begin{equation}
  \label{eq:gavoret}
  G^{N}(K) \sim - G^{A}(K)\sim\frac{\rho^{0}}{\rho}
  \frac{m c^2}{\omega^2+c^2 k^2},\;\;K\to0,
\end{equation}
where $c$ is the velocity of the Goldstone modes,
$\rho^{0}$ is the condensate density and $\rho$ 
denotes the boson density. Note that expression (\ref{eq:gavoret}) 
is an {\it exact} asymptotic result. However, as was mentioned by Gavoret 
and Nozi\`{e}res themselves, these asymptotic formulas 
were initially obtained by 
%simply 
omitting divergent terms which appear in the diagrams for 
both self-energies. 
Gavoret and Nozi\`{e}res further assumed
that
$\Sigma^{A}(0)\neq0$, which contradicts the 
Nepomnyashchy relation. \cite{Nepomnyashchy1975} 
Eq.~(\ref{eq:gavoret}) was later rederived in a formalism which 
is free of IR divergences and is consistent with $\Sigma^{A}(0)=0$, see
Ref.~[\onlinecite{Nepomnyashchy1975}]. We sketch here briefly the arguments which 
lead to the result $\Sigma^{A}(0)=0$.

The anomalous self-energy can be 
represented as a sum of one-particle irreducible skeleton diagrams,
some of which are regular and some IR divergent,
\begin{figure}[htb]
  \centering
  \includegraphics[height=4.cm]{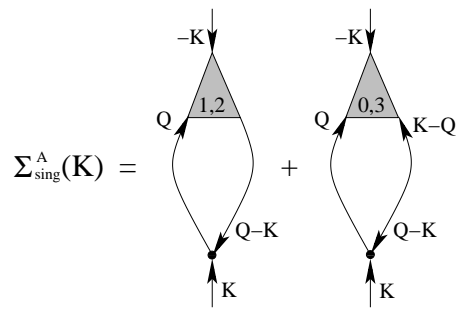} 
  \caption{Diagrams giving rise to the IR-divergent contribution~(\ref{Nep1})
 to the anomalous self-energy.
 Black dots denote the bare vertices while
the shaded triangles correspond to fully renormalized vertices.}
  \label{fig:Nep2}
\end{figure}
\begin{equation}
  \label{Nep1} 
  \Sigma^A(K) = \Sigma^{A}_{\rm reg}(K)+\Sigma^{A}_{\rm sing}(K),
\end{equation}
where the first term on the right-hand side contains terms with a 
single closed Green's function. These diagrams are
regular for $K\to0$ and $D>1$. 
The second term in Eq.~(\ref{Nep1}) is IR divergent and 
arises from 
the diagrams shown in Fig.~\ref{fig:Nep2} which have the analytic form
\begin{widetext}
  \begin{eqnarray}
    \Sigma^{A}_{\rm sing}(K) =  
    2u_{\Lambda_0}
    \sqrt{\rho^{0}}\intop_{Q}\left\{{\Gamma}^{(1,2)}(Q,K,Q-K)G^{N}(Q)G^{N}(Q-K)
      + {\Gamma}^{(0,3)}(Q-K;Q,-K)G^{N}(Q)G^{A}(K-Q)\right\},
 \nonumber
 \\
 & &  \label{eq:Nep2} 
  \end{eqnarray}
\end{widetext}
where ${\Gamma}^{(1,2)}$ and ${\Gamma}^{(0,3)}$ are the exact
irreducible vertices with two in- and one outgoing, and
with three ingoing legs, respectively. The first factor 
$u_{\Lambda_0}\sqrt{\rho^0}$ in Eq.~(\ref{eq:Nep2})
is the value of the bare vertex  
${\Gamma}_{\Lambda_0}^{(1,2)}$.
%Note that they are different from the three-point vertices used in the perturbative  expansion, which are given by equation~(\ref{eq:Gamma12}).
% It can be seen by a simple dimensional analysis that three-legged vertices of the skeleton expansion should scale as $\tilde\Gamma^{(3)}\propto E^2/\rho$, while perturbative vertices as $\Gamma^{(3)}\propto E/\sqrt{\rho}$, where $E$ and $\rho$ denote some abstract energy and density, respectively. 
%The diagrammatic representation of equation~(\ref{eq:Nep2}) is shown in 
%Figure~\ref{fig:Nep2}.  
%Equation~(\ref{eq:Nep2}) is exact up to an irrelevant constant factor. 
In the limit $K=0$ and for small internal momenta $Q$ we employ  
the asymptotic formulas~(\ref{eq:gavoret}) and obtain
%\begin{widetext}
  \begin{eqnarray}
     \nonumber
    \Sigma^{A}(0) &=& \left[{\Gamma}^{(1,2)}(0,0,0)-{\Gamma}^{(0,3)}(0,0,0)\right] \\
&\times& K_D \intop_0^{\Lambda^{}_0}~dq~q^{D-1}\int\frac{d\omega}{2\pi}
    \frac{2u_{\Lambda_0}\sqrt{\rho^0}}{(\omega^2+c^2q^2)^2},\;\;
\label{eq:Nep3}
  \end{eqnarray}
%\end{widetext}
where the upper limit of the integration over momenta $q$ is irrelevant 
since we are only interested in the IR behavior 
of the integral. 
The expression 
in the brackets on the right-hand side of  equation~(\ref{eq:Nep3}) 
can be further simplified by using the 
Ward identity \cite{Nepomnyashchy1975} 
\begin{equation}
  \label{eq:Nep4} 
  % \frac{1}{2}\left[
  {\Gamma}^{(1,2)}(0,0,0)-{\Gamma}^{(0,3)}(0,0,0)
  % \right] = %\frac{1}{\rho^0}
  =\frac{\Sigma^{A}(0)}{\sqrt{\rho^0}},
\end{equation}
% where the proportionality factor depends only on the 
which, upon insertion of  Eq.~(\ref{eq:Nep4}) into 
Eq.~(\ref{eq:Nep3}), yields an exact self-consistent equation for the
 anomalous self-energy. Since the integral on the right-hand side of equation~(\ref{eq:Nep3}) diverges for $D\le 3$, 
the only finite solution of this equation is
\begin{equation}
\label{eq:Nep5} 
\Sigma^{A}(0) = 0.
\end{equation}

\section{Analytic approach to the $z=2$ regime of the derivative expansion}
\label{sec:HardCore}

Here we want to investigate the flow of the interaction constant
$u_\Lambda$ in the regime where $z_\Lambda=2$, i.e. the initial
part of the renormalization group flow. This regime yields
results which are analogous to the renormalization of the
$T$-matrix in both $D=2$ and $D=3$. The two-dimensional
situation is especially interesting since the scattering length
vanishes in $D=2$. We work out below the relevant scale which replaces
the scattering length and recover the results first
discussed by
Schick \cite{Schick1971} who investigated the hard-core Bose gas in 
two dimensions using diagrammatic techniques. 
In $D=2$ the vanishing  of $u_\Lambda$
for $\Lambda\to 0$ in the initial $z=2$ regime is 
logarithmic in $\Lambda/\Lambda_0$, 
which reflects the vanishing of the 
$s$-wave scattering length.\cite{Fisher1988} 
%This requires a modified 
%perturbative approach in terms of the new small parameter $1/\ln(\rho d^2)$, 
%where $\rho$ denotes the density of bosons and $d$ is related to the size of the 
%hard core disc.  
In the $z=2$ regime, it suffices to investigate the flow of 
the interaction, while setting all remaining coupling parameters
of the derivative expansion to their initial 
values. Performing the integration over frequencies we obtain from
Eq.~(\ref{eq:uFlow})
\begin{eqnarray}
\hspace{-.3cm}
  \partial^{}_\Lambda u^{}_\Lambda & \approx & u^2_\Lambda 
\frac{\Lambda^{D+1}K_D}{2mD} \frac{\epsilon^3_\Lambda-5\epsilon^{2}_\Lambda
\Delta^{}_\Lambda+\epsilon^{}_\Lambda\Delta^{2}_\Lambda
+3\Delta^{3}_\Lambda}{E^5_\Lambda},\;\;\;
\label{eq:int}
\end{eqnarray}
where $E^{}_\Lambda=\sqrt{\epsilon^{}_\Lambda(\epsilon^{}_\Lambda
+2\Delta^{}_\Lambda)}$
% and $\kappa^{}_{D}=K^{}_D/D$.
and $\Delta_\Lambda=\rho_\Lambda^0 u_\Lambda$.
In the $z=2$ regime, the Bogoliubov spectrum approaches the free particle 
dispersion, i.~e. $\epsilon^{}_\Lambda\gg\Delta^{}_\Lambda$. 
In this case we may simplify Eq.~(\ref{eq:int}) as follows
\begin{equation}
\label{eq:HC:IntSimp}
\partial^{}_\Lambda u^{-1}_\Lambda \approx - 2m (K_D/D) \Lambda^{D-3} \, .
\end{equation}
For $D>2$  the
solution of Eq.~(\ref{eq:HC:IntSimp}) is of the form
\begin{equation}
\label{eq:HC:SolDneq2} 
u^{}_\Lambda \approx \frac{u^{}_{\Lambda_0}}
{1+\displaystyle\frac{2mu^{}_{\Lambda_0}K_D}{D(D-2)}
\left(\Lambda^{D-2}_0-\Lambda^{D-2}\right)} \, .
\end{equation}
For $\Lambda \to 0$ the interaction flows to a finite value 
which for $u_{\Lambda_0} \Lambda_0^{D-2}\gg 1$ is given by
\begin{equation}
  \label{eq:HC:uPhysDgg2}
  u^{}_{\Lambda\to0} \approx \frac{u^{}_{\Lambda_0}}
  {1+\displaystyle\frac{2mu^{}_{\Lambda_0}K_D}{D(D-2)}\Lambda^{D-2}_0} 
\approx \frac{(D-2)a^{D-2}}{2m (2 \pi)^{D-2}\kappa^{}_D},
\end{equation}
where we related the UV-cutoff $\Lambda^{}_0\approx 2 \pi/a $ 
to the inverse $s-$wave scattering 
length $a$. In $D=3$ we obtain 
\begin{equation}
u_{\Lambda \to 0} \approx \frac{3\pi}{2 m}a, 
\end{equation}
which is smaller than the scattering theory prediction for the $T-$matrix 
$8\pi a/m$~(see for instance Ref.~[\onlinecite{Shi1997}]). 
However, this can be accounted for by re-adjusting the UV-cutoff to 
$\Lambda^{}_0 = 3\pi/8a$. According to Eq.~(\ref{eq:HC:SolDneq2}), 
the $s$-wave scattering length is finite in all dimensions above two. 

% On the other hand, expression (\ref{eq:HC:SolDneq2}) approaches zero as 
% $\Lambda\to0$ in dimensions below two. This behavior is power-law-like, i.~e.
% \begin{equation}
% \label{eq:HC:uPhysDll2}
% u^{}_{\ell\to\infty} \approx \frac{2-D}{2m\kappa^{}_D}\Lambda^{2-D}.
% \end{equation}
% Hence, the $s-$wave scattering length vanishes in all dimensions $D<2$. From Eq.~(\ref{eq:HC:uPhysDll2}) follows 
% an interesting consequence for the anomalous self-energy at zero frequency and momenta, which thus vanishes also in the $z=2-$regime. 

In $D=2$, the solution of Eq.~(\ref{eq:HC:IntSimp}) has the form
\begin{equation}
  \label{eq:intFH}
  u^{}_\Lambda  \approx \frac{u^{}_{\Lambda_0}}{1+\displaystyle
    \frac{m u^{}_{\Lambda_0}}{2\pi}\ln\left(\frac{\Lambda_0}{\Lambda}\right)}
  \, .
\end{equation}
In the regime 
\begin{equation}
1\gg \frac{m u^{}_{\Lambda_0}}{2\pi}\ln\left(\frac{\Lambda_0}{\Lambda}\right) 
\end{equation}
the interaction does not flow and the Bogoliubov theory remains correct. 
The opposite case
\begin{equation}
  1\ll\frac{mu^{}_{\Lambda_0}}{2\pi} \ln\left(\frac{\Lambda_0}{\Lambda}\right)
\end{equation}
corresponds to the limit of hard-core bosons, where the flow does not depend 
on the initial value of the interaction. In this case we may write
\begin{equation}
\label{eq:HCB}
u^{}_\Lambda\approx \frac{2\pi}{m \ln\left(\frac{\Lambda_0}{\Lambda}\right)},
\end{equation}
which reproduces the result obtained by Fisher and 
Hohenberg.\cite{Fisher1988} 
If we now define the crossover scale where the scaling behavior 
of the Bogoliubov spectrum changes as 
$\epsilon^{}_{\Lambda_c} \approx  2\rho^{0}u^{}_{\Lambda_c}$, 
we find, setting up to logarithmic accuracy 
$\ln(\Lambda_0/\Lambda_c)^2\approx 1$, 
\begin{equation}
  \label{eq:crovscaleII}
  \Lambda^2_c \approx 16 \pi\rho.
\end{equation}
Hence, the corresponding energy crossover scale is given by
\begin{equation}
  \label{eq:chempot}
  \Delta_{\Lambda_c}\approx \rho u^{}_{\Lambda_c}\approx-
\frac{8 \pi \rho}{m\ln{(\rho^0 d^2)}}, 
\end{equation}
which is Schick's crossover scale for hard-core 
bosons,\cite{Schick1971}
where $d$ is of the order of the diameter of the  hard-core disc and we chose
$\Lambda_0 = \sqrt{16\pi/ d^2}$. The velocity of the Goldstone mode in the hard-core regime becomes
\begin{equation}
  \label{eq:GoldVelHB} 
  c =  \sqrt{\frac{\Delta^{}_{\Lambda_c}}{2m}}
\approx\sqrt{-\frac{4\pi\rho}{m^2\ln{(\rho d^2)}}} \, .
\end{equation}
Thus, the FRG flow equations qualitatively describe the hard-core limit. 
%A new small parameter for the perturbative expansion in $D\leqslant2$ emerges naturally from these equations. 

\section{Asymptotic behavior of the derivative expansion}
\label{sec:GizbScale}

Here we derive the asymptotic $\Lambda\to 0$ behavior of the flow equations
obtained within the derivative expansion, Eqs.~(\ref{eq:rhoFlow},
\ref{eq:uFlow}, \ref{eq:YFlow}-\ref{eq:VFlow}), and derive from them
the Ginzburg scale $k_G$ which is characteristic for this regime.
This regime differs from the one discussed in Appendix~\ref{sec:HardCore}
in that now the dynamical exponent is $z_\Lambda=1$. 
The non-perturbative character
of the Bose gas is manifest in this regime.

For $\Lambda \to 0$ we may replace all coupling parameters which remain
finite in this limit by their fixed point values,
i.e. $Z_\Lambda \approx Z$, $V_\Lambda\approx V$, 
$\rho^0_\Lambda \approx \rho^{0}$. We may further approximate
$\eta^z_\Lambda \approx 0$ since $\eta_\Lambda^z$ plays no important role at $T=0$.
We are thus left with the flows of $u^{}_\Lambda$ and $Y^{}_\Lambda$. 
Assuming $|\omega|\leqslant c\Lambda$, where $c$ denotes the 
renormalized velocity of the Goldstone mode defined in Eq.~(\ref{eq:GoldVel}), the
small $\Lambda$ behavior of the denominator in the 
propagators~(\ref{eq:GNapprox}) and 
(\ref{eq:GAapprox}) is \cite{Dupuis2007}
\begin{eqnarray}
  \nonumber
  &&Y^2_\Lambda \omega^2+(\epsilon^{}_\Lambda+V^{}_\Lambda\omega^2)
  (2\Delta^{}_\Lambda+\epsilon^{}_\Lambda+V^{}_\Lambda\omega^2)  \\
  \label{eq:AsPropKK}
  &\approx& 2\Delta^{}_\Lambda \left(\epsilon^{}_\Lambda+V\omega^2
  \right) = 2V\Delta^{}_\Lambda(\omega^2+c^2\Lambda^2). 
\end{eqnarray}
For small $\Lambda$, the leading contribution in the numerator of 
Eq.~(\ref{eq:uFlow}) arises from the term  $4\Delta^3_\Lambda$ in 
$c^{(u)}_0$ defined in Eq.~(\ref{eq:cu0}), since all remaining 
terms contain additional powers of
$\epsilon^{}_\Lambda\propto \Lambda^2$. 
The integration over frequencies can be carried out and the flow equation 
of the interaction Eq.~(\ref{eq:uFlow}) simplifies to 
\begin{equation}
  \label{eq:IntFlowIII}
  \partial^{}_\Lambda u^{}_\Lambda  \approx \frac{u^2_\Lambda}{A^{}_D}  
  \Lambda^{D-4},
\end{equation}
where 
\begin{equation}
\label{eq:eval:A_D}
A^{}_D=\frac{4 D}{3mk^{}_c K_D}\left(\frac{\rho^0}{\rho^s} 
\right)^2 = \frac{A^\prime_D}{mk^{}_c}.
\end{equation}
Here $k^{}_c=2mc$ and $\rho^s=Z\rho^0$ is the superfluid density
which at $T=0$ coincides with the density of bosons \cite{Dupuis2007} and
\begin{equation}
\label{eq:eval:A^pr_D}
A^\prime_D = \frac{4 D}{3 K_D}\left(\frac{\rho^0}{\rho^s} \right)^2.
\end{equation}
One has to distinguish between the $D<3$ and $D=3$ cases. 
In the first case, the general solution of Eq.~(\ref{eq:IntFlowIII}) 
can be found in the form 
\begin{equation}
\label{eq:int:Dneq3} 
u_\Lambda \approx \frac{u^{}_{c}\Lambda^{\epsilon}}
{\Lambda^{\epsilon}+\displaystyle\frac{mk^{}_cu^{}_{0}}
{\epsilon A^{\prime}_D}
\left[1-\left(\frac{\Lambda}{\Lambda^{}_c}\right)^\epsilon\right]},
\end{equation}
where $\epsilon=3-D$. Here, we introduced the scale $\Lambda_c$ and
$u_{\Lambda_c}=u_c$. The scale $\Lambda_c$ is the characteristic scale
for the boundary of the perturbative $z=2$ regime and can be approximated as 
$\Lambda_c\approx k_c$.
At this scale, $u_\Lambda$ has already been renormalized by the 
RG flow through the perturbative $z=2$ regime, loosely
defined by $\Lambda_c \lesssim \Lambda \lesssim \Lambda_0$. This in general can lead
to a substantial suppression of $u_c$ from its bare value
$u_{\Lambda_0}$. However, for weak coupling, for an order of magnitude
estimate it is sufficient to assume $u_c\approx u_{\Lambda_0}$.
In any event, the flow of the interaction for $\Lambda \ll \Lambda_c$
in $D<3$ neither depends on the chosen initial interaction $u^{}_c$,
  nor on the starting value for the cutoff-parameter $\Lambda_c$, but 
only on the fully renormalized quantities entering the momentum  $k^{}_c$ 
and the factor $A^\prime_D$. 
%Therefore, we can identify starting value of the
% cutoff-parameter with the UV-cutoff $\Lambda^{}_0$ and initial value for 
%the interaction with the bare interaction $u^{}_{0}$. For $\Lambda \ll \Lambda_G$
%the flow is however 
For $\epsilon>0$, i.~e. for $D<3$, 
the asymptotic expression for $u_\Lambda$ is in this limit given by
\begin{equation}
  \label{eq:ueps}
  u^{}_{\Lambda \to 0} \sim \frac{\epsilon A^\prime_D}{ mk^{}_c}\Lambda^\epsilon \, .
\end{equation}
For $D<3$ the crossover from the $z=2$ to the 
$z=1$ regime is governed by the generalized Ginzburg scale 
$k_G$, see Refs.~[\onlinecite{SHK2008,Kreisel2007,Dupuis2009}] %The flow of the interaction for $\Lambda<k^{}_G$ does not depend on the initial condition directly. 
which can be read off from Eq.~(\ref{eq:int:Dneq3})
%\begin{equation}
%\label{eq:int:GinzScDef} 
%k^{\epsilon}_G\approx\frac{mk^{}_cu^{}_{0}}{\epsilon A^{\prime}_D}\left[1-\left(\frac{k^{}_G}{\Lambda^{}_0}\right)^\epsilon\right].
%\end{equation}
%Resolving Eq.~(\ref{eq:int:GinzScDef}) with respect to $k^{}_G$, we obtain
\begin{equation}
\label{eq:int:GinzSc} 
k^{}_G \approx 
%\left[\frac{\displaystyle\frac{mk^{}_c u^{}_{G}}{\epsilon A^{\prime}_D}}
%{1+\displaystyle\frac{mk^{}_c u^{}_{0}}
%{\epsilon A^{\prime}_D\Lambda^{\epsilon}_0}} \right]^{\frac{1}{\epsilon}}
\left[\frac{mk^{}_cu^{}_{0}}
{\epsilon A^{\prime}_D}\right]^{\frac{1}{\epsilon}} \, .
\end{equation}
For the weakly interacting Bose gas we have
$k_c\approx 2m c_0$ and $\rho^s\approx\rho^{0}$. Then, $k_G$ 
may be rewritten in a form similar to the one which was obtained by 
Castellani {\it et al.} \cite{Pistolesi2004} 
and Kreisel {\it et al.}, \cite{Kreisel2007}
\begin{equation}
  \label{eq:int:GinzScII} 
  k^{}_G \approx \left(\frac{3K_D}{16\epsilon D}\right)^{\frac{1}
{\epsilon}} \left(\frac{k^D_{c}}
{\rho^{}}\right)^{\frac{1}{\epsilon}}  k_c \, .
\end{equation}
%The corresponding flowing parameter $\ell^{}_G$ is then found to be
%\begin{equation}
%\ell^{}_G\approx\frac{1}{\epsilon} \ln\left(\Lambda^{\epsilon}_0\frac{\epsilon A^{\prime}_D} {mk^{}_cu^{}_{0}}\right). 
%\end{equation}

\begin{figure}[t]
\includegraphics[height=5cm]{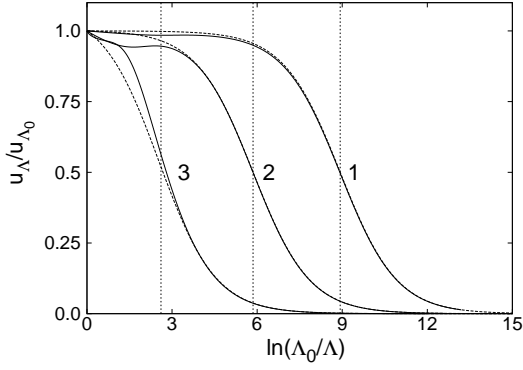}
\caption{Comparison of the full numerical solution of the flow equation 
for the interaction $u_\Lambda$ as obtained from Eqs.~(\ref{eq:rhoFlow},
\ref{eq:uFlow}, \ref{eq:YFlow}-\ref{eq:VFlow})
(solid lines) in $D=2$  with the approximate solution 
Eq.~(\ref{eq:int:Dneq3}) (dashed lines) for different values of the bare parameters. 
Solutions 1 are calculated with 
$\tilde{u}^{}_0=0.1$, $\tilde{\mu}^{}_0=0.001$ which implies 
$k^{}_c/k^{}_G\approx340$. Solutions 2 were calculated with $\tilde{u}^{}_0=0.8$ 
and $\tilde{\mu}^{}_0=0.08$ with $k^{}_c/k^{}_G\approx45$. Solutions 3 are 
obtained with $\tilde{u}^{}_0=15$ and $\tilde{\mu}^{}_0=0.15$ with 
$k^{}_c/k^{}_G\approx 5$. The vertical dotted lines show positions of the 
Ginzburg scale for each solution.}
\label{fig:IntComp}
\end{figure}

In Fig.~\ref{fig:IntComp} we  compare of the full solution 
of the flow of $u_\Lambda$, obtained from Eqs.~(\ref{eq:rhoFlow},
\ref{eq:uFlow}, \ref{eq:YFlow}-\ref{eq:VFlow}),  with the 
approximate solution Eq.~(\ref{eq:int:Dneq3}) for $D=2$ and
different initial conditions. 
While solutions marked with numbers 1 and 2 in Fig.~\ref{fig:IntComp} correspond 
to a weakly interacting system (see the legend), i.e. the ratio $c/c^{}_0\approx1$, 
where $c$ represents the Goldstone mode velocity calculated from 
Eq.~(\ref{eq:GoldVel}) and 
$c^{}_0=\sqrt{\rho^0 u_{\Lambda_0}/m}$ 
corresponds to the mean-field 
result for the Goldstone mode velocity, the solution marked 
with the number 3 corresponds to a more strongly interacting system with 
$c/c^{}_0\approx0.669$. It is evident that the approximate solution 
Eq.~(\ref{eq:int:Dneq3}) always yields a very satisfactory approximation for 
$\Lambda\leqslant\Lambda^{}_c$.

In $D=3$, the only non-trivial solution of Eq.~(\ref{eq:IntFlowIII}) reads
\begin{equation}
\label{eq:int3Dsol}
u^{}_\Lambda \approx  \frac{u^{}_{c}}{1+\displaystyle\frac{mk^{}_cu^{}_{c}}
{A^\prime_3}\ln{\displaystyle\left(\frac{\Lambda^{}_{c}}{\Lambda^{}}\right)}},
\end{equation}
which yields the Ginzburg scale
\begin{equation}
\label{eq:CrOvSc}
k^{}_G \approx \Lambda^{}_c\exp\left(-\frac{A^{\prime}_3}{mk^{}_cu^{}_{c}}\right),
\end{equation}
where $\Lambda_c \approx k_c$.
In the weakly interacting regime, this expression reduces to
\begin{equation}
  \label{eq:CrOvScII}
  k^{}_G \approx k_c\exp\left(-\frac{8\pi^2\rho^0}{k^3_{c^{}}}\right) \, ,
\end{equation}
which is also consistent with the result reported 
in Refs.~[\onlinecite{Pistolesi2004}, \onlinecite{Kreisel2007}].

Analogously, from Eq.~(\ref{eq:YFlow}) we obtain,
\begin{equation}
  \label{eq:EqSFlowII}
  \partial^{}_\Lambda Y^{}_\Lambda \approx \frac{u^{}_\Lambda Y^{}_\Lambda}
{A^{}_D} \Lambda^{D-4} \, .
\end{equation}
Taking in account the initial condition $Y^{}_c\approx 1$, the
solution of Eq.~(\ref{eq:EqSFlowII}) reads 
\begin{equation}
  \label{eq:YY:Dneq3} 
  Y^{}_\Lambda \approx \frac{\Lambda^{\epsilon}}{\Lambda^{\epsilon}+\displaystyle\frac{mk^{}_cu^{}_{c}}{\epsilon A^{\prime}_D}\left[1-\left(\frac{\Lambda}
{\Lambda^{}_c}\right)^\epsilon\right]},
\end{equation}
in $D<3$, and 
\begin{equation}
  \label{eq:YY3Dsol}
  Y^{}_\Lambda \approx  \frac{1}{1+\displaystyle\frac{mk^{}_cu^{}_{0}}
    {A^\prime_3}\ln{\displaystyle\left(\frac{\Lambda^{}_{c}}
        {\Lambda^{}}\right)}} \, .
\end{equation}
in $D=3$. Hence, the generalized Ginzburg scale is also characteristic for the flow of the coupling parameter $Y^{}_\Lambda$.

\section{Asymptotic behavior of propagators in the infrared limit}
\label{sec:AsymGreen}

We here extend the derivative expansion analysis of Sec.~\ref{sec:derivative}
to include also an expansion of $u_{\Lambda} ( K )$
to second order in momentum and frequency. 
We thus approximate  
%Eq.~(\ref{eq:approxsigmaa}) by the expression
\begin{equation}
  \label{eq:Intassym}
  u_\Lambda(K) \approx u^{}_\Lambda+\alpha_\Lambda 
    \epsilon_{k}+\beta_\Lambda \omega^2 \, . 
\end{equation}
As mentioned in Sec.~\ref{sec:derivative}, the coupling function 
$u_\Lambda(K)$ is expected
to become non-analytic for $\Lambda \to 0$ and this can be understood already
partially from the scaling dimensions of $\alpha_\Lambda$ and $\beta_\Lambda$, 
%We find the scaling dimensions of coupling parameters $\alpha^{}_\ell$ and $\beta^{}_\ell$  to be
\begin{subequations}
  \begin{eqnarray}
    \label{eq:SDalpha}
    [\alpha_\Lambda] & = & 2-D-z_\Lambda,\\
    \label{eq:SDbeta}
    [\beta_\Lambda] & = & 4-D-3z_\Lambda. 
  \end{eqnarray}
\end{subequations}
In the Goldstone regime we have $z_\Lambda=1$ which for $D=2$ yields 
$[\alpha_\Lambda]=[\beta_\Lambda]=-1$, indicating that both $\alpha_\Lambda$
and $\beta_\Lambda$ diverge as $\Lambda^{-1}$ for small $\Lambda$.
This suggests that for small $K$ the leading order behavior of 
$u_{\Lambda \to 0}(K)$ should be proportional to $k$ or $|\omega|$,
which is consistent with a $\sqrt{\omega^2 + c^2 k^2}$ behavior which 
we find from the full frequency and momentum dependent calculation.
In $D=3$, the dimensional analysis predicts 
$[\alpha_\Lambda]=[\beta_\Lambda]=-2$.
This is suggestive of a logarithmic dependence on $|\omega|$ and $k$ which is
indeed correct.

The flow equations for $\alpha^{}_\Lambda$ and $\beta^{}_\Lambda$ can
be obtained from 
\begin{subequations}
  \begin{eqnarray} 
    \label{eq:FRGalfa}
    \partial^{}_\Lambda\alpha^{}_\Lambda & = & -\frac{\alpha^{}_\Lambda}
    {\rho^{0}_\Lambda}\partial^{}_\Lambda\rho^{0}_\Lambda+\left.
      \frac{m}{\rho^{0}_\Lambda}\partial^{}_\Lambda\left(\frac{\partial^2}{\partial k^2}\Sigma^{A}_\Lambda({\bm k},0)\right|_{k=0}\right),\\
\label{eq:FRGbeta}
\partial^{}_\Lambda\beta^{}_\Lambda & = & -\frac{\beta^{}_\Lambda}{\rho^{0}_\Lambda}\partial^{}_\Lambda\rho^{0}_\Lambda+\left.\frac{1}{2\rho^{0}_\Lambda}\partial^{}_\Lambda\left(\frac{\partial^{2}}{\partial\omega^2}\Sigma^{A}_\Lambda(0,\omega)\right|_{\omega = 0}\right) ,\; \;\;\;\;\;\;\;\;
\end{eqnarray}
\end{subequations}
where we employ a non-self-consistent approach to calculate the flows, i.e.
the flows are calculated in exactly the same approximation
as the full frequency and momentum dependent calculation presented in
Sec.~\ref{sec:Spectral}. 
The initial conditions of $\alpha^{}_\Lambda$ and $\beta^{}_\Lambda$ 
are $\alpha^{}_{\Lambda_0}=0$ and $\beta^{}_{\Lambda_0}=0$. 

\begin{figure}[t]
  \centering
  \includegraphics[height=5cm]{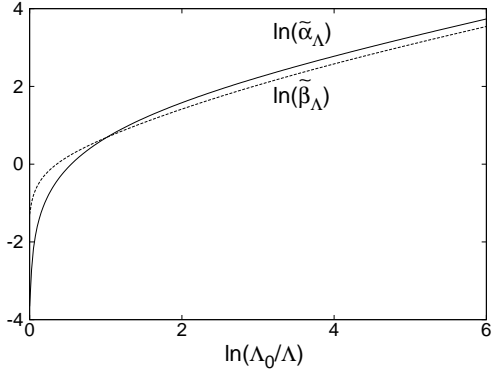}
  \caption{
%    THIS PLOT SHOULD BE ON A LOG LOG SCALE, OTHERWISE ONE CANNOT SEE THE
%    NATURE OF THE DIVERGENCE
%    RG flow of the dimensionless parameters 
    $\tilde\alpha^{}_\Lambda=\Lambda^D_0\alpha^{}_\Lambda$ and 
    $\tilde\beta^{}_\Lambda~=~2m\Lambda^{D-2}_0\beta^{}_\Lambda$, 
    calculated in $D=2$ for the initial values of the dimensionless 
    chemical potential $\tilde\mu^{}_0=2m\mu\Lambda^{-2}_0=0.4$ and 
    dimensionless interaction $\tilde{u}^{}_0=2m u_{\Lambda_0}\Lambda_0^{D-2}=4$.} 
  \label{fig:AB}
\end{figure}

The analytic expression for the flows is rather lengthy and can be found
in Ref.~[\onlinecite{Diss}]. 
The flow of the parameters $\alpha^{}_\Lambda$ and $\beta^{}_\Lambda$ for
$D=2$ is shown in Fig.~\ref{fig:AB}. 
The divergence of both parameters 
can be understood by keeping only the
leading order terms of their flow equations in the limit $\Lambda\to 0$. 
In this case, both flow equations reduce to
\begin{equation}
  \label{eq:ABsimple} 
  \partial^{}_\Lambda
  \alpha^{}_\Lambda \propto
  \partial^{}_\Lambda \beta^{}_\Lambda 
  \propto u^2_\Lambda \Lambda^{D-6} \, .
\end{equation}
For $D<3$ we have from Eq.~(\ref{eq:ueps}) 
$u^{}_\Lambda\propto\Lambda^{3-D}$ and thus the
solution of Eqs.~(\ref{eq:ABsimple}) has the  following 
small $\Lambda$ behavior,
\begin{equation}
  \alpha^{}_\Lambda \propto
  \beta^{}_\Lambda
  \propto  \Lambda^{1-D},
\end{equation}
which agrees with the dimensional estimate for $z_\Lambda=1$. 
In $D=3$, the small $\Lambda$ behavior 
$u_\Lambda$ is proportional to $\ln(\Lambda/\Lambda_0)$ [see
Eq.~(\ref{eq:int3Dsol})], so that for small $\Lambda$ Eq.~(\ref{eq:ABsimple}) 
leads to
\begin{equation}
\alpha^{}_\Lambda \propto
\beta^{}_\Lambda
\propto
\intop_{\Lambda} d\Lambda^\prime \frac{1}{(\Lambda^{\prime})^{3} 
\ln^2 (\Lambda_0/\Lambda^\prime)}\propto\frac{1}
{\Lambda^2\ln(\Lambda_0/\Lambda)}\,\, ,
\end{equation}
where we have kept only the leading term. 
The upper integration boundary is not relevant for this 
estimate.
Hence, using $\Lambda \sim \sqrt{\omega^2+c^2 k^2}$,
we obtain the analytical behavior of the physical  
anomalous self-energy in the low-energy limit in $1<D<3$,
\begin{eqnarray}
  \nonumber
  \Sigma^{A}(K) &\propto& 
  (\omega^2+c^2 k^2)^{\epsilon/2}\, .
\label{eq:AsAnSE}
\end{eqnarray}
%where $\epsilon=3-D$ and
%\begin{eqnarray}
%  \label{eq:eval:C_D}
%  C^{}_D &\approx &
%  \frac{5}{6}\frac{\epsilon^2}{2-\epsilon}\frac{A^\prime_D\rho^{}_\star }
%  {mk^{}_c c^{\epsilon}},
%\end{eqnarray}
%$k^{}_c=2mc$, and $A^{}_D$ and $A^{\prime}_D$ are defined in Eqs.~(\ref{eq:eval:A_D}) and (\ref{eq:eval:A^pr_D}).

In $D=3$, the leading behavior of the anomalous self-energy is logarithmic,
\begin{eqnarray}
\nonumber
\Sigma^{A}(K) &\propto& \frac{1}%C^{}_3}
{\displaystyle\ln\left(\frac{c\Lambda^{}_0}{\sqrt{\omega^2+c^2{k}^2}}\right)} \, .
\end{eqnarray}
%where 
%\begin{equation}
%  \label{eq:eval:C_3}
%  C^{}_3 = -\frac{5}{9}\frac{A^\prime_3\rho^{}_\star}{mk^{}_c}.
%\end{equation}

We are now in the position to recover the low-energy behavior of the 
Green's functions. By taking  Eqs.~(\ref{eq:approxsigman}, {\ref{eq:approxsigmaa},
\ref{eq:Denom2}, \ref{eq:Intassym}) into account, 
one finds
\begin{eqnarray}
\label{eq:G11trunc} \nonumber
G^{N}_\Lambda(K) & = &\frac{1}{{\cal D}^{}_\Lambda(K)}
\left[\Delta^{}_\Lambda+ iY^{}_\Lambda\omega + (Z_\Lambda^{-1} 
+\rho^{0}_\Lambda\alpha^{}_\Lambda) \epsilon^{}_{  k}\right.\\
&+&\left. R^{}_\Lambda({  k}) + (V^{}_\Lambda + 
\rho^{0}_\Lambda\beta^{}_\Lambda)\omega^2\right],\,\;\;\;\;\\
\label{eq:G20trunc} 
G^{A}_\Lambda(K) & = & -\frac{1}{{\cal D}^{}_\Lambda(K)}
\left(\Delta^{}_\Lambda+\rho^{0}_\Lambda\alpha^{}_\Lambda\epsilon^{}_{  k}
+\rho^{0}_\Lambda\beta^{}_\Lambda\omega^2\right),\;\;\;\;\;\;
\\
\nonumber
{\cal D}^{}_\Lambda(K) & = & Y^2_\Lambda\omega^2+(\epsilon^{}_{  k}/Z_\Lambda
+R^{}_\Lambda({  k})+V^{}_\Lambda\omega^2)\\
\nonumber
&\times  & [(Z_\Lambda^{-1}+2\rho^{0}_\Lambda\alpha^{}_\Lambda)
\epsilon^{}_{  k} +R^{}_\Lambda({  k}) \\
\label{eq:DenFull} 
&&+(V^{}_\Lambda+2\rho^{0}_\Lambda\beta^{}_\Lambda)\omega^2].
\end{eqnarray}

The divergence of the parameters $\alpha^{}_\Lambda$ and $\beta^{}_\Lambda$ 
enables us to expand Eqs.~(\ref{eq:G11trunc}, \ref{eq:G20trunc}) in 
powers of the small quantity $\alpha^{-1}_\Lambda$ 
under the assumption $\beta^{}_\Lambda/\alpha^{}_\Lambda\to const$. 
For momenta $k \lesssim \Lambda$, 
we obtain to leading order of the expansion,
\begin{eqnarray}
  \nonumber
  G^{N}( K) & = & -G^{A}(K) = \frac{1}{2V
[\omega^2+(2mZ V)^{-1}k^2]}\\
  \label{eq:PropExact}
  & = & \frac{m\rho^{0} c^2}{\rho}\frac{1}{\omega^2+c^2k^2},
\end{eqnarray}
where $\rho=\rho^0/Z$, and the velocity of the Goldstone mode is 
\begin{equation}
  \label{eq:GoldVel2}
  c = \frac{1}{\sqrt{2 m V Z}} \, ,
\end{equation}
see Eq.~(\ref{eq:GoldVel}). Eq.~(\ref{eq:PropExact}) is exactly 
the Gavoret and Nozi\`{e}res result\cite{Gavoret1964} 
given in Eq.~(\ref{eq:gavoret}). We conclude
that our non-self-consistent approach from Sec.~\ref{sec:Spectral}
both correctly describes the non-analytic behavior of the self-energies
and yields the correct asymptotic behavior of the propagators. Note also
that in the asymptotic result (\ref{eq:PropExact}) the scale $k_G$
and all of the non-analytic behavior of the self-energies is completely
absent and it remains an open problem whether and how the non-perturbative
scale $k_G$ shows up in the spectrum or damping of quasiparticles.
The scale does however emerge quite dramatically in the so-called
longitudinal Green's function which is defined and discussed e.g. 
in Refs.~[\onlinecite{Pistolesi2004,Dupuis2009,Dupuis2009b}]. 
While for usual Bose
systems the longitudinal Green's function is not accessible to experiments,
in antiferromagnetic materials, subject to strong magnetic fields close
to the saturation field strength, it is possible to directly probe
the spectral properties of the longitudinal Green's function,\cite{Kreisel2007}
which is related to the longitudinal spin structure factor of the underlying
spin model.

\end{document}